\documentclass[amsmath,amssymb,11pt]{article}
\usepackage{jheppub2}
\pdfoutput=1
\usepackage{graphicx,epsfig}
\usepackage{float}
\usepackage{subfig}
\usepackage{subfloat}
\usepackage[utf8]{inputenc}
\usepackage{hyperref}
\usepackage{caption}
\usepackage{color}

\newcommand{\gl}{\ell_\star}

\newcommand*{\affmark}[1][*]{\textsuperscript{#1}}

\newcommand{\beq}{\begin{equation}}

\newcommand{\eeq}{\end{equation}}

 \newcommand{\be}{\begin{equation}}
 \newcommand{\ee}{\end{equation}}
 \newcommand{\bea}{\begin{eqnarray}}
 \newcommand{\eea}{\end{eqnarray}}

\usepackage{tikz}
\usetikzlibrary{positioning}
\usetikzlibrary{intersections}
\usetikzlibrary{fadings} 
\usetikzlibrary{arrows.meta} 
\usetikzlibrary{arrows}

\tikzfading[name=fade out,
inner color=transparent!0,
outer color=transparent!100]

\definecolor{cherryblossompink}{rgb}{1.0, 0.72, 0.77}
\definecolor{lightblue}{rgb}{0.68, 0.85, 0.9}

\usetikzlibrary{decorations.pathmorphing}
\usetikzlibrary{decorations.pathreplacing,decorations.markings}

\usetikzlibrary{backgrounds,automata}

\title{Nucleation of charged quantum de-Sitter$_{3}$ black holes}

\author{Ana Climent,\affmark[1]}
\author{Robie A. Hennigar,\affmark[1]}
\author{Emanuele Panella,\affmark[2]}
\author{and Andrew Svesko\affmark[3]}

\affiliation{\affmark[1]Departament de Física Quàntica i Astrofísica and
  Institut de Ciències del Cosmos,\\
 Universitat de Barcelona, 08028 Barcelona, Spain\\
 \affmark[2]Department of Physics and Astronomy, University College London,\\
Gower Street, London, WC1E 6BT, United Kingdom\\
\affmark[3]Department of Mathematics, King’s College London,
Strand, London, WC2R 2LS, United Kingdom}

\emailAdd{anacliment@icc.ub.edu}
\emailAdd{robie.hennigar@icc.ub.edu}
\emailAdd{emanuele.panella.21@ucl.ac.uk}
\emailAdd{andrew.svesko@kcl.ac.uk}

\abstract{We construct charged, static black holes in three-dimensional de Sitter (dS$_{3}$) space that exactly account for semi-classical backreaction effects due to quantum conformal matter. This is accomplished using braneworld holography, where an accelerating, electrically charged anti-de Sitter$_{4}$ black hole localizes on a Randall-Sundrum end-of-the-world brane. Absent of backreaction, the black hole disappears, leaving a chemical conical defect. The ``quantum'' black hole has a physical parameter space characterized by a shark-fin diagram, with extremal, Nariai, and ultracold limits. We give a detailed analysis of the horizon thermodynamics, where we find the heat capacity of charged and neutral dS$_{3}$ black holes features Schottky peaks. In particular, for a specific temperature range, charged quantum black holes  behave as thermal systems with a finite number of energy levels available to their underlying microscopic degrees of freedom, beyond which many energy levels become available. Finally, we compute the probability of nucleating quantum dS black holes. Our work gives a first step to study quantum matter backreaction effects on dS black hole decay.

}

\begin{document}

\maketitle

\section{Introduction} \label{sec:intro}


Black holes in four-dimensional de Sitter (dS) space are incredibly rich. Due to exponential inflation in the future owed to a positive cosmological constant, a static observer is restricted to a patch of the whole spacetime; their view obstructed by a cosmological horizon. Thus, to a static patch observer, dS black holes have at least two visible horizons.  Consequently, their thermal \cite{Gibbons:1977mu}  and subsequent microscopic description is more subtle than their flat or anti-de Sitter relatives \cite{Banks:2000fe,Witten:2001kn,Spradlin:2001pw, Goheer:2002vf,Banihashemi:2022jys,Banihashemi:2022htw}. 
Further, when charged, dS black hole decay has been conjectured to bound masses of elementary particles \cite{Montero:2019ekk}, placing constraints on the Higgs potential and possibly explaining the relation between neutrino mass and vacuum energy \cite{Montero:2021otb}. De Sitter black holes, then, are of broad appeal, from a theoretical and phenomenological standpoint. Notably absent, however, is a consistent description of how quantum matter modifies de Sitter black hole physics. 

For a macroscopic observer, incorporating quantum matter effects in a classical background amounts to solving the semi-classical Einstein equations,
\beq G_{\mu\nu}+\Lambda g_{\mu\nu}=8\pi G_{\text{N}}\langle T_{\mu\nu}\rangle\;.\label{eq:scEin}\eeq
On the left is the Einstein tensor for a classical spacetime $g_{\mu\nu}$ with cosmological
constant $\Lambda$, while on the right is the expectation value of the (renormalized) stress-tensor of quantum fields. To date, no generic four-dimensional black hole solution to (\ref{eq:scEin}) exists, even when the ``backreaction'' on the geometry due to quantum fields -- and vice versa -- is treated perturbatively. Progress can be made by descending to one  dimension lower. In fact, semi-classical backreaction has a most dramatic effect in $(2+1)$-dimensional asymptotically flat or dS space: the generation of black holes where there were none before \cite{Emparan:2022ijy,Panella:2023lsi}. 

To wit, there are no black hole solutions to three-dimensional Einstein gravity with $\Lambda\geq0$.\footnote{Black holes in dS$_{3}$ have been uncovered in topologically massive theories \cite{Nutku:1993eb,Anninos:2009jt}, and ``new massive gravity'' \cite{deBuyl:2013ega}, including perturbative backreaction effects \cite{Chernicoff:2024dll}.} Rather, Schwarzschild- and Kerr-(dS$_{3}$) geometries describe a conical defect with (possibly) a single cosmological horizon and no black hole horizon \cite{Deser:1983nh,Deser:1983tn,Klemm:2002ir}. Backreaction due to quantum fields produces a Casimir effect with a negative energy density so as to  produce a gravitationally attractive effect \cite{Souradeep:1992ia,Soleng:1993yh}, paving the way for black hole formation. A simple example of this is to consider dS$_{3}$ Einstein gravity and a conformally coupled scalar field in a Hartle-Hawking-like state. The renormalized stress-tensor of the scalar field in Schwarzschild-dS$_{3}$ has the form \cite{Emparan:2022ijy}
(see \cite{Panella:2023lsi} for Kerr-dS$_{3}$)
\beq \langle T^{\mu}_{\;\nu}\rangle=\frac{\hbar F(M)}{8\pi r^{3}}\;\text{diag}(1,1,-2)\;, \label{eq:Tmunugen}\eeq
for radial coordinate $r$, and some positive function $F(M)$ of the point mass generating the conical defect. Substituting the stress-tensor (\ref{eq:Tmunugen}) into the semi-classical field equations (\ref{eq:scEin}), the linear order perturbative correction  modifies the $tt$-component of the metric such that the blackening factor shifts $H(r)\to H(r)-\frac{2L_{\text{P}}F(M)}{r}$, with Planck length $L_{\text{P}}=\hbar G_{3}$. It is not known how to continue the algorithm to include higher-order corrections analytically.

In this article we construct the first electrically charged dS$_{3}$ black hole \emph{exactly} incorporating all orders of semi-classical backreaction. 
Namely, a charged ``quantum'' black hole in dS$_{3}$.
Our approach, following \cite{Emparan:2002px,Emparan:2020znc,Emparan:2022ijy,Panella:2023lsi,Climent:2024nuj}, is by way of braneworld holography \cite{deHaro:2000wj}. In this framework, the anti-de Sitter/conformal field theory (AdS/CFT) correspondence is applied to scenarios where a portion of a bulk AdS$_{d+1}$ spacetime, including its conformal boundary where the CFT$_{d}$ resides, is removed by a $d$-dimensional end-of-the-world (ETW) Randall-Sundrum \cite{Randall:1999vf,Randall:1999ee} or Karch-Randall \cite{Karch:2000ct,Karch:2000gx} braneworld.  The geometry of the ETW brane is dynamical, having an induced theory of gravity coupled to a CFT$_{d}$ in its large-$c$ central charge, planar diagram limit.
From the perspective of a macroscropic observer confined to the brane, the induced theory may be interpreted as a semi-classical theory of gravity, where higher-derivative corrections incorporate backreaction due to the CFT. It was thus conjectured that any bulk black hole which localizes on the brane is as an \emph{exact} solution to the induced semi-classical field equations, i.e., a quantum black hole \cite{Emparan:2002px} (see \cite{Panella:2024sor} for a review). 

The charged quantum dS$_{3}$ black hole we construct has the line element (\ref{eq:qsdsc})
\beq ds^{2}=-H(r)dt^{2}+H^{-1}(r)dr^{2}+r^{2}d\phi^{2}\;,\quad H(r)=1-8\mathcal{G}_{3}M-\frac{r^{2}}{R_{3}^{2}}-\frac{\ell F}{r}+\frac{\ell^{2}Z}{r^{2}}\;.\label{eq:qsdscintro}\eeq
Here, $\mathcal{G}_{3}$ is the ``renormalized'' Newton constant, $R_{3}$ is the dS$_{3}$ length scale, and $F$ and $Z$ are functions dependent on the mass $M$ and electric charge $Q$ of the black hole. From the brane perspective, the parameter $\ell$ controls the strength of backreaction, and
$\ell\approx 2cL_{\text{P}}$. This shows when backreaction vanishes, both the $1/r$ and $1/r^{2}$ terms turn off (leaving Schwarzschild-dS$_{3}$), demonstrating the attractive gravitational potential and the effects of electric charge arise solely due to backreaction of a CFT$_{3}$ coupled to a gauge field. 
Notably, moreover, since $c L_{\text{P}}\gg L_{\text{P}}$, the new terms in (\ref{eq:qsdscintro}) are a bona fide semi-classical effect, yielding a black hole horizon far exceeding the Planck scale where quantum gravitational effects become relevant. Similar to quantum Kerr-dS$_{3}$ \cite{Panella:2023lsi}, the charged dS$_{3}$ black hole has a physical parameter space characterized by a ``shark fin'' diagram with extremal, Nariai, and ultracold limits.

Backreaction leads to an enriched thermal description of the quantum black hole. In particular, the Bekenstein-Hawking entropy \cite{Bekenstein:1973ur,Hawking:1974sw} of the bulk black hole localizing on the ETW brane is, via holographic duality, identified with the three-dimensional \emph{generalized} entropy \cite{Bekenstein:1974ax}, accounting for both the gravitational entropy (including the infinite tower of higher-derivative terms) and the von Neumann entropy of the CFT$_{3}$. Further, akin to their classical four-dimensional counterparts \cite{Johnson:2019ayc,Dinsmore:2019elr}, we show the heat capacity of quantum dS$_{3}$ black holes have (inverted) Schottky peaks. Specifically, we show both classical and quantum charged black holes feature a maximum temperature where the heat capacity diverges, beyond which many energy levels become available. This signals that charged quantum black holes, for a particular range of temperature, serve as thermal systems with a finite number of energy levels available to its underlying microscopic degrees of freedom, after which many energy levels are again available, indicating a large gap in the energy spectrum. 

Finally, we compute the probability of nucleating  charged quantum dS$_{3}$ black holes using the (constrained) instanton method \cite{Cotler:2020lxj,Draper:2022xzl,Morvan:2022ybp,Morvan:2022aon}. 
The probability for an arbitrary mass and charged black hole is expressed as an integral over dimensionless mass and charge parameters weighted by the deficit of generalized entropies between (quantum) dS$_{3}$ and the charged black hole, (\ref{eq:nucrateqbhgen}). We compute the probability numerically, finding  it decreases as the strength of backreaction increases (at fixed central charge). This is a consequence of the fact that large backreaction on the brane coincides with freezing out gravity, whilst removing influential non-perturbative contributions. Our work here can be viewed as a  first step to study the effects of backreaction of quantum matter on dS black hole decay and its phenomenological implications.

The remainder of this article is organized as follows. In Section \ref{app:ds3_defects} we analyze the physics of charged defects in dS$_{3}$ using AdS$_{4}$/CFT$_{3}$. The bulk spacetime is (a double Wick rotation of) Reissner-Nordstr{\"o}m-AdS$_4$ whose boundary geometry is a charged defect dS$_{3}$.  We compute the holographic stress-tensor  and entanglement entropy in the defect background. We construct charged quantum black holes in dS$_{3}$ and analyze their geometric features in Section \ref{sec:GeomKdS3}, and horizon thermodynamics in Section \ref{sec:thermo}. In Section \ref{sec:schotpeaks} we investigate the Schottky behavior of charged classical and quantum de Sitter black holes. Section \ref{sec:entdefs} is devoted to computing the nucleation of charged quantum-dS$_{3}$ black holes. We summarize our findings and discuss future avenues worth exploring in Section \ref{sec:disc}. We also include two appendices. Appendix \ref{app:range_bulk} details the physical parameter range of the constructed quantum black holes. Appendix \ref{app:nucleation} exactly computes the nucleation of conical defects in dS$_{3}$, and provides the first numerical investigation of the nucleation of (classical) charged dS$_{4}$ black holes.

\section{Charged defects in dS$_3$}
\label{app:ds3_defects}

One of the powerful applications of the AdS/CFT duality is to study properties of strongly coupled quantum fields on curved spacetimes. Here we are interested in understanding the physics of chemical defects coupled to quantum fields in a de Sitter universe and, later, their quantum backreaction. We begin by constructing bulk duals to describe strongly coupled quantum fields on the background
\be
\label{eq:bdry_geom}
ds^2 = - \left(\alpha^2 - \frac{r^2}{R_3^2}\right)dt^2 + \frac{dr^2}{\alpha^2 - \frac{r^2}{R_3^2}} + r^2 d\phi^2 
\ee
supplemented with the chemical potential
\be
\label{eq:defect_pot}
A = \frac{a}{r} dt \, .
\ee
Here $R_3$ denotes the cosmological length scale of the de Sitter universe, while $\alpha$ characterizes whether the geometry has a conical singularity ($\alpha < 1$), a conical surfeit ($\alpha > 1$), or is regular ($\alpha = 1$). The chemical potential is characterized by a single parameter $a$ which corresponds to a marginal deformation of the CFT. The chemical potential is singular at the origin and should be regarded to arise from a pointlike electric charge.

To construct the bulk dual to this boundary geometry, here and throughout this work our focus will be on Einstein-Maxwell theory, for which we take the following action
\be \label{eq:bulk_action}
I_{\text{bulk}} = \frac{1}{16 \pi G_4} \int d^4x \sqrt{-g} \left[\frac{6}{\ell_{4}^2} + \hat{R} - \frac{\gl^2}{4} F_{\mu\nu}F^{\mu\nu} \right] \, .
\ee
The length scale $\gl$ measures the coupling between the gravitational and electromagnetic sectors of the theory,
\be 
\gl^2 = \frac{16 \pi G_4}{g_\star^2} \, ,
\ee
for a (dimensionless) gauge coupling constant $g_\star$. 

\subsection{Bulk solution and holographic data}

The Einstein-Maxwell equations admit the following solution that has as its boundary geometry the chemical defect in dS$_3$ \eqref{eq:bdry_geom} with~\eqref{eq:defect_pot},
\begin{align}\label{eq:bdry_defect}
ds^2 &= \frac{\alpha^2 \rho^2}{R^2} \left[-\left(\alpha^2 - \frac{R^2}{R_3^2} \right) dt^2 + \frac{dR^2}{\alpha^2 - \frac{R^2}{R_3^2}} + \frac{R^2 \ell_4^2 }{\rho^2} F(\rho) d\phi^2 \right] + \frac{d\rho^2}{F(\rho)} \, ,
\\
A &= \frac{a}{R} dt \, ,
\end{align}
where
\be 
F(\rho) = \frac{\rho^2}{\ell_4^2} - 1 - \frac{2 G_4 m}{\rho} - \frac{e^2}{\rho^2} \, , \quad e  = - \frac{a \gl }{2 \alpha^2} \, .
\ee
Here we have distinguished the coordinate $R$ in the bulk from the coordinate $r$ that appears in the boundary geometry~\eqref{eq:bdry_geom} for reasons that will be clarified momentarily.

The constants $m$ and $e$ (or equivalently $m$ and $a$) are integration constants of the bulk solution. These parameters are not completely free, and must be constrained such that the bulk does not contain a naked singularity. If the metric function $F(\rho)$ has no zeroes, then the bulk spacetime contains a naked timelike curvature singularity. On the other hand, if the metric function has a zero such that $F(\rho_0) = 0$ with $\rho_0 > 0$, then the bulk geometry pinches off and the curvature singularity is not part of the spacetime. For the spacetime to pinch off smoothly at $\rho = \rho_0$ requires
\be 
\alpha = \frac{2}{\ell_4} \frac{1}{| F'(\rho_0)|} \, .
\label{eq:alphacond}\ee
Here we shall restrict ourselves to parameter choices such that the spacetime pinches off smoothly with $\alpha$ determined in (\ref{eq:alphacond}). We will characterize more completely the allowed parameter ranges below. 

The above solution is a double Wick rotation of the familiar magnetically charged Reissner-Nordstr{\"o}m-AdS$_4$ metric. To see this, start with the Reissner-Nordstr{\"o}m-AdS$_4$ metric
\begin{align}
    ds^2 &= - F(\rho) dT^2 + \frac{d\rho^2}{F(\rho)} + \rho^2 d\sigma^2 + \rho^2 \sinh^2\sigma d\Phi^2 \, ,
    \\
    A &= - \frac{2 g}{\gl} \cosh \sigma d\Phi  \, ,
    \quad 
    F(\rho) = \frac{\rho^2}{\ell_4^2} - 1 - \frac{2 G_4 m}{\rho} + \frac{g^2}{\rho^2}
\end{align}
and perform the following complex coordinate transformation
\be 
T = i \alpha \ell_4 \phi \, ,\quad \Phi = i \frac{\alpha t}{R_3} \, , \quad \sinh^2 \sigma  = \frac{\alpha^2 R_3^2}{R^2} - 1 \, ,\quad g = - i e \, .
\ee
This precisely gives the bulk metric (\ref{eq:bdry_defect}). 

Let us now construct the holographic dictionary for the bulk solution~\eqref{eq:bdry_defect}. By transforming the bulk coordinates to 
\be 
\rho = \frac{\ell_4 r}{z} - \frac{z \ell_4}{r}\left(\frac{1}{2} - \frac{r^2}{4 \alpha^2 R_3^2} \right) + \frac{m z^2}{3 r^2} + \cdots  \, , \quad R = r + \left(\alpha^2 - \frac{r^2}{R_3^2}\right)\frac{ z^2 }{2 r \alpha^2} + \cdots  \, .
\ee
the metric can be cast into Fefferman-Graham form,
\begin{align} 
ds^2 &= \frac{\ell_4^2}{z^2} \left[dz^2 + \gamma_{ij}(z, x) dx^{i} dx^{j}  \right] \, ,
\\
\gamma_{ij}(x,z) &= \gamma_{ij}^{(0)}(x) + z^2 \gamma_{ij}^{(2)}(x) + z^3 \gamma_{ij}^{(3)}(x) + z^4 \gamma_{ij}^{(4)}(x) + \dots \, .
\end{align}
The boundary metric and gauge field  obtained in the $z \to 0$ limit are precisely those given in eqs.~\eqref{eq:bdry_geom} and~\eqref{eq:defect_pot}.

The current response is vanishing, i.e., $\left< J^{i} \right> = 0$, but the point charge is immersed in a background electric field, with field strength
\be 
F_{tr} = \frac{2 \alpha^2 e}{\gl r } \, ,
\ee
given by the boundary value of the field strength tensor.  From the third-order correction in the Fefferman-Graham expansion, we can read off the holographic stress tensor expectation value $\left<T_{ij} \right>  \equiv 3 \ell_4^2/(16 \pi G_4) \gamma^{(3)}_{ij}$, giving  
\be 
\left<T_j^{i} \right> = \frac{c}{8\pi} \frac{\alpha^2 m}{ r^3 \ell_4} \, {\rm diag} \left(1,1,-2 \right) \, .
\label{eq:expvalTmunudefect}\ee
Note the physical effect of the defect. The boundary geometry has a cosmological horizon at $r = \alpha R_3$ with proper circumference $2\pi \alpha R_3$. Thus if the defect is a conical deficit, the size of the cosmological horizon is decreased. This is physically analogous to adding positive Killing energy to the static patch of higher dimensional de Sitter space in the form of black holes. However, an important detail in the present case is that due to the presence of charge and chemical potential it is possible to have conical deficits even when the bulk parameter $m$ (and hence the energy of the CFT state) is negative. On the other hand, conical excesses lead to a cosmological horizon with larger proper circumference. This is analogous to what happens when negative Killing energy is added to the static patch of higher dimensional de Sitter space~\cite{Gibbons:1977mu, Banihashemi:2022htw}. 

\subsection{Defect entropy}

We have introduced a class of geometries that describe a charged defect immersed in a CFT plasma on a non-dynamical conical de Sitter background. A natural question is to what extent this defect is entangled with its surroundings. This problem can be addressed via the Ryu-Takayanagi prescription \cite{Ryu:2006bv,Ryu:2006ef}.

Consider an entangling surface at a constant value of the boundary coordinate $r = R_{\rm RT}$ enclosing the defect. It is easy to see that this surface extends into the bulk trivially as an extremal surface. We can compute the entanglement entropy by computing the area of this bulk surface,
\be 
S_{\rm EE} = \frac{2 \pi \alpha \ell_4}{4 G_4} \int_{\rho_0}^{\rho_{\rm max}} d \rho  = \frac{ \pi \alpha \ell_4}{2 G_4} \left(\rho_{\rm max} - \rho_0 \right) \, .
\ee
Here we have introduced a UV cutoff $\rho_{\rm max}$ to regulate the usual divergence that occurs near the AdS boundary. We can obtain a finite result by subtracting the case when the boundary is pure dS$_3$ without a defect present, which corresponds to $\rho_0 = \ell_4$ and $Q = 0$ in the bulk. In performing the subtraction, one must be careful to match the the proper size of the $\phi$ circle at the cutoff surface. Taking this detail into account, we obtain the finite result
\be 
\Delta S_{\rm EE} \equiv S_{\rm defect} - S_{{\rm dS}} =  \frac{\pi \ell_4^2}{2 G_4} \left(1 - \frac{\alpha}{\lambda} \right) \, .
\ee  
In the last line we introduced the dimensionless parameter $\lambda = \ell_4/\rho_0$. The factor $\ell_4^2/G_4$ can be swapped for the CFT central charge, $c = \ell_4^2/G_4$. 

\begin{figure}[t!]
\centering 
\includegraphics[width=0.75\textwidth]{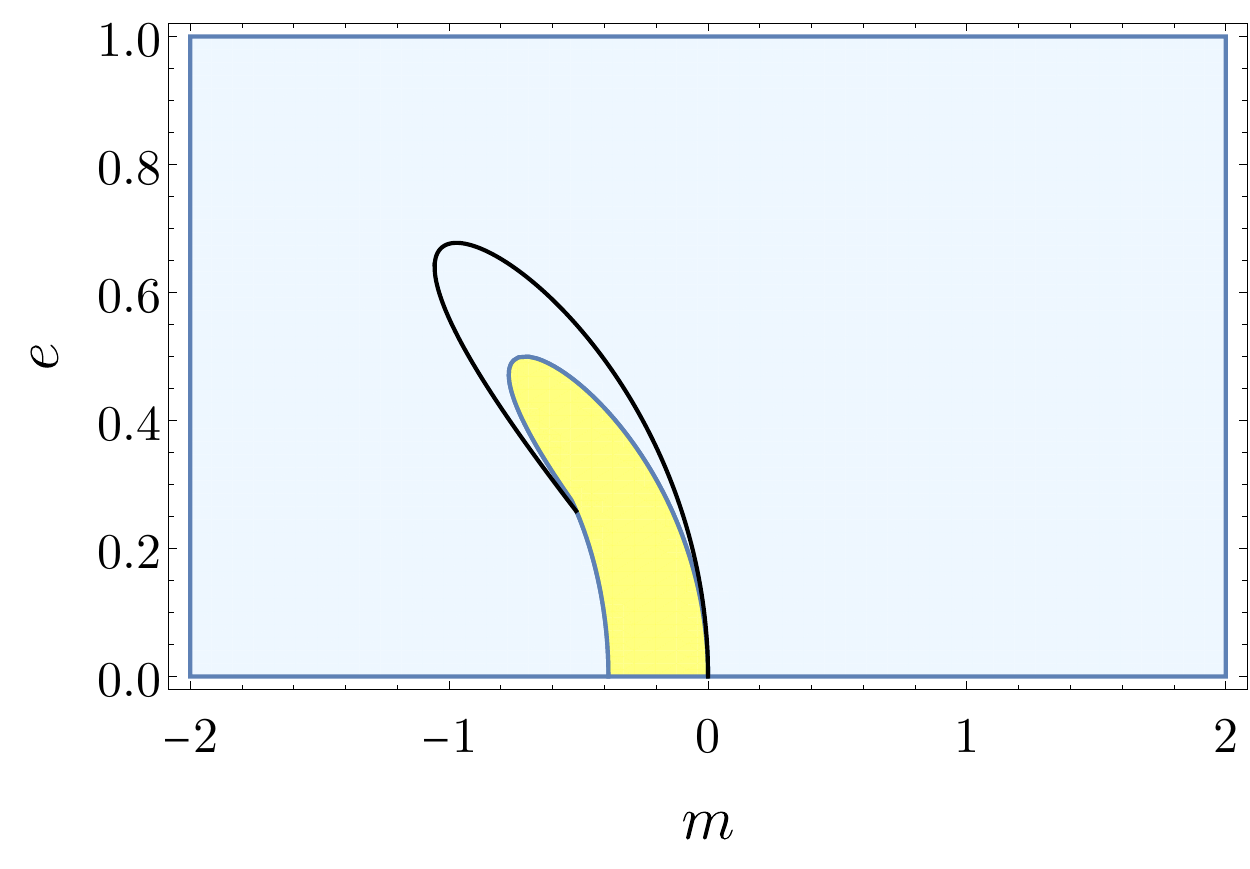}
\caption{Sign of the defect entropy as a function of the bulk parameters $(m, e)$. In the blue-shaded region the defect entropy is positive, while in the yellow shaded region it is negative. The black curve corresponds to the geometries with $\alpha = 1$ and is included to make clear that the defect entropy only becomes negative for a subset of the conical surfeits. Units are such that $\ell_4 = 1$.}
\label{fig:ent_banana}
\end{figure}

Notice the final result $\Delta S_{\text{EE}}$ is independent of the entangling surface; the same result is obtained for any constant-$r$ surface enclosing the defect. A similar phenomenon was obtained in~\cite{Horowitz:2014gva}, where the authors considered a charged defect in a Minkowski background. The fact $\Delta S_{\text{EE}}$ does not depend on the entangling surface led those authors to interpret the result as measuring the \textit{defect entropy}, i.e.,~the entanglement between the defect and its surroundings~\cite{Affleck:1991tk,Jensen:2013lxa}. 

We show in Figure~\ref{fig:ent_banana} the sign of $\Delta S_{\rm EE}$ as a function of the bulk parameters $(m, e)$. For the greater part of the parameter space the defect entropy is positive. The defect entropy is negative only for a small region of parameter space (indicated by the yellow banana region), which corresponds to a subset of the conical surfeits. The solid black line in the plot indicates regular geometries with $\alpha = 1$, highlighting it is a \textit{proper subset} of the conical surfeits that have negative defect entropy.

An important question concerns whether or not the defect entropy calculated above receives strong quantum corrections. We shall not fully address this problem here, but present preliminary evidence that it may. Consider again the bulk geometry~\eqref{eq:defect_pot} describing the charged defect in dS$_3$. Transform the bulk coordinate $R = -\alpha^2/r$ such that the spacetime metric reads
\be 
ds^2 = \rho^2 \left[- \left(\frac{r^2}{R_3^2} - \alpha^2\right) dt^2 + \frac{dr^2}{\frac{r^2}{R_3^2} - \alpha^2}\right] + \alpha^2 \ell_4^2 F(\rho) d\phi^2 + \frac{d\rho^2}{F(\rho)} \, .
\ee
It is now immediate to recognize the $(t, r)$ sector of the spacetime is AdS$_2$-Rindler. As a result of recent studies, e.g.~\cite{Iliesiu:2020qvm, Emparan:2023ypa}, it is known that geometries with an AdS$_2$ factor --- most notably, extremal black holes --- receive strong quantum gravitational corrections that can drive the entropy to zero. It is plausible similar quantum gravitational corrections are relevant here.

\section{Charged quantum-$\text{dS}_{3}$ black holes} \label{sec:GeomKdS3}

Here we construct an exact three-dimensional quantum black hole with charge. The gauge field and geometry can be understood as the result of the letting the quantum conformal fields backreact on the geometries studied in section~\ref{app:ds3_defects}. This is accomplished via braneworld holography, where it has been conjectured  classical solutions to the bulk theory obeying brane boundary conditions may be interpreted as solutions to the induced semi-classical theory on the brane \cite{Emparan:2002px}. Our analysis below follows the analogous constructions of neutral quantum de Sitter black holes \cite{Emparan:2022ijy,Panella:2023lsi}, and the charged quantum BTZ solution \cite{Climent:2024nuj}.


\subsection{Bulk and brane geometry}

\subsection*{Bulk geometry}


We consider Einstein-Maxwell theory with a negative cosmological constant in a four-dimensional bulk. The action and conventions are the same as those in eq.~\eqref{eq:bulk_action}. A particular solution of interest to (\ref{eq:bulk_action}) is the charged $\text{AdS}_{4}$ C-metric, whose line element in Boyer-Lindquist-like coordinates is
\beq ds^{2}=\frac{\ell^{2}}{(\ell+xr)^{2}}\left[-H(r)dt^{2}+\frac{dr^{2}}{H(r)}+r^{2}\left(\frac{dx^{2}}{G(x)}+G(x)d\phi^{2}\right)\right]\;,\label{eq:AdS4Ccoord}\eeq
with metric functions $H(r)$ and $G(x)$
\beq H(r)= 1-\frac{r^{2}}{R_{3}^{2}}-\frac{\mu\ell}{r}+\frac{q^{2}\ell^{2}}{r^{2}}\;,\qquad G(x)=1-x^{2}-\mu x^{3}-q^{2}x^{4}\;.\label{eq:Hrfunc}\eeq
Our conventions primarily follow \cite{Climent:2024nuj}, except we set the discrete parameter $\kappa=+1$ and $\ell_{3}^{2}=-R_{3}^{2}$ (or $\ell_{3}=iR_{3}$) such that the Randall-Sundrum brane we eventually introduce has a $\text{dS}_{3}$ radius $R_{3}$.\footnote{The cases $\kappa=0$ or $\kappa=-1$ exclude the possibility of a $\text{dS}_{3}$ brane, since the roots of $H(r)$ do not represent a cosmological horizon in those cases.} The C-metric may be interpreted as a single or pair of uniformly accelerating black holes (in our case a pair) \cite{Kinnersley:1970zw},  where the real, positive  parameter $\ell$ is equal to the (inverse) acceleration.  Meanwhile, $\mu>0$ is interpreted as a mass parameter of the four-dimensional black hole. The $\text{AdS}_{4}$ length scale $\ell_{4}$ is related to $R_{3}$ and $\ell$ via
\beq \ell_{4}^{-2}=\ell^{-2}\left[1-\left(\frac{\ell}{R_{3}}\right)^{2}\right]\;.\label{eq:L4bulk}\eeq
For $\ell_{4}^{2}>0$ such that the bulk cosmological constant is negative, we require $R_{3}^{2}>\ell^{2}$. We can think of $R_{3}$ as the dS$_{3}$ length scale on the brane. The $U(1)$ gauge field $A_{\mu}$ takes the form
\be \label{eq:gauge_Cmetric}
    A_{\mu}dx^{\mu} = \dfrac{2q \ell }{\gl} \left(\frac{1}{r_i} - \frac{1}{r} \right) dt\, , 
\ee
with electric charge parameter $q$, such that the only non-vanishing component of the Maxwell tensor is $F_{rt}=2q\ell/(\ell_{\star}r^{2})$.\footnote{In principle one could include magnetic charge, such that the gauge field has an additional component $A_{\phi}=\dfrac{2 \ell }{\gl}g(x-x_{1})$, for real parameter $x_{1}$.} Here $r_{i}$ refers to the largest root of $H(r_{i})=0$. 

As for the neutral black hole, real roots $\{x_{i}\}$ of the metric function $G(x)$ correspond to symmetry axes of the Killing vector $\partial^{a}_{\phi}$. Each zero produces a conical singularity which distorts the black hole horizon caused by a cosmic string attached to the horizon that pulls the black hole away from the center of AdS$_{4}$ toward the conformal boundary. A conical singularity at, say, $x=x_{1}$, may be removed via the identification 
\beq \phi\sim\phi+2\pi\Delta \;,\quad \Delta=\frac{2}{|G'(x_{1})|}=\frac{2x_1}{|-3+ x_{1}^{2}-q^{2}x_{1}^{4}|}\;,\label{eq:conedefidsds}\eeq
where we treat $\mu$ as a derived parameter by solving $G(x_{1})=0$, 
\beq \mu=\frac{1- x_{1}^{2}-q^{2}x_{1}^{4}}{x_{1}^{3}}\;.\label{eq:muqtdef}\eeq
Once the period of $\phi$ has been fixed, the conical singularities at poles $x_{i}\neq x_{1}$ remain. As such, we restrict ourselves to the parameter range $0\leq x\leq x_{1}$, where $x_{1}$ is the smallest positive real root. Having included $q$, the parameter $x_{1}$ belongs to a different parameter range than the static or rotating cases \cite{Emparan:2022ijy,Panella:2023lsi} (see Appendix \ref{app:range_bulk}). Crucially, $\Delta(x)$ is not monotonic  in $0\leq x \leq 1$, for $\mu<0$. 
 
\subsection*{Brane geometry}

A key geometric feature of the C-metric (\ref{eq:AdS4Ccoord}) is that the $x=0$ hypersurface is totally umbilic, i.e., the extrinsic curvature $K_{ij}$ is proportional to the induced metric $h_{ij}$ of the hypersurface at $x=0$. Consequently, a brane placed at $x=0$ will automatically satisfy the Israel-junction conditions, 
\beq ([K_{ij}]-h_{ij}[K])=-8\pi G_{4}S_{ij}\;,\label{eq:Israeljuncconds}\eeq
where $[K_{ij}]\equiv K_{ij}^{+}-K_{ij}^{-}$ with `$+$' and `$-$' refer to the spacetime on either side of the brane, $K=h^{ij}K_{ij}$, and $S_{ij}$ is the brane stress-tensor. In our case we will have a double-sided brane obeying $K_{ij}^{+}=-K_{ij}^{-}\equiv K_{ij}$. 

\begin{figure}[t!]
\begin{center}
\includegraphics[width=.2\textwidth, height=.32\textwidth]{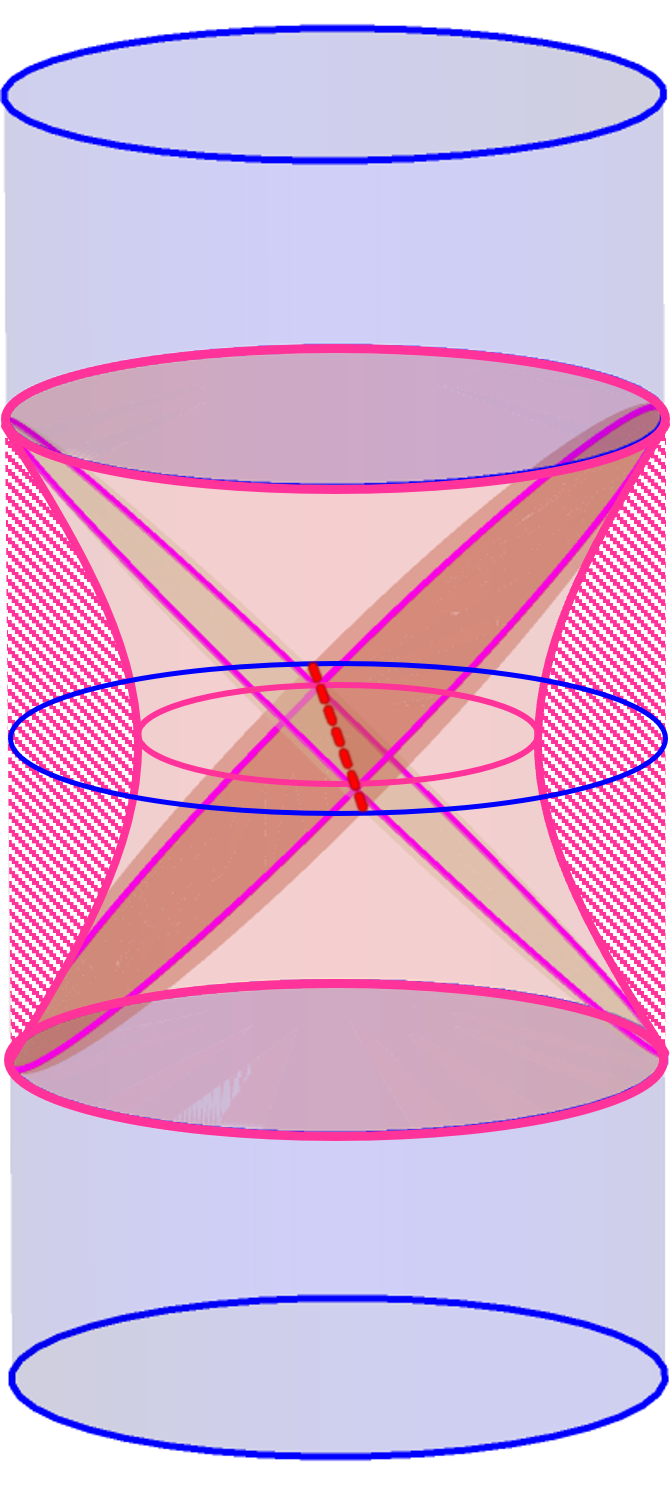} $\quad\quad\quad\quad$ \includegraphics[width=.3\textwidth]{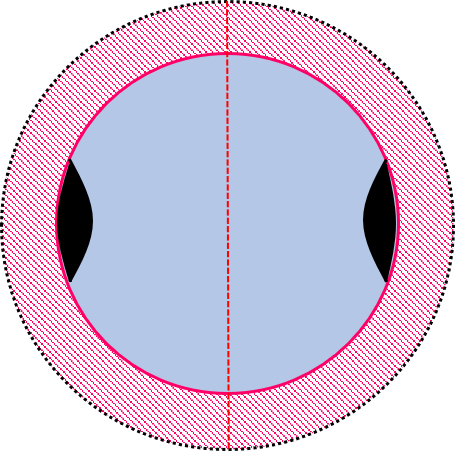}
\end{center}
\caption{\small \textbf{Left:} Bulk $\text{AdS}_{4}$ with a dS$_3$ end of the world brane. The brane is represented as a hyperboloid. The bulk region up to the brane ($x<0$, dashed magenta region) is excluded. To complete the construction, a second copy is glued along $x=0$. Cosmological horizons on the dS brane correspond to bulk acceleration horizons intersecting the brane (red dashed line). \textbf{Right:} Constant $t$-time slice of a single $\text{AdS}_{4}$ cylinder with a de Sitter brane (thick red circle) containing black holes. The coordinates cover only half of the disk, containing only a single black hole and cosmological horizon (dashed red line); the other half is obtained via an appropriate analytic continuation.}
\label{fig:dSbranebhs} 
\end{figure}

Following the construction of \cite{Emparan:1999wa,Emparan:1999fd}, we place an end-of-the-world (ETW) Randall-Sundrum brane $\mathcal{B}$ \cite{Randall:1999vf} at $x=0$. Assuming a purely tensional action for convenience,
\beq I_{\text{brane}}=-\tau\int_{\mathcal{B}}d^{3}x\sqrt{-h}\;,\label{eq:braneact}\eeq
the junction conditions (\ref{eq:Israeljuncconds}) set the tension $\tau$ to be
\beq \tau=\frac{1}{2\pi G_{4}\ell}\;.\label{eq:branetens}\eeq
Thus, tuning the tension corresponds to changing the position of the brane. As an ETW brane, the bulk space is cutoff at $x=0$, and we keep the $x>0$ portion of the bulk. See Figure \ref{fig:dSbranebhs} for an illustration. To complete the space we glue a second copy of this construction along the brane at $x=0$ such that the brane is two-sided and the bulk is $\mathbb{Z}_{2}$ symmetric.  Due to this spacetime surgery, the remaining conical singularities at $x_{i}\neq x_{1}$ are removed from the completed bulk geometry. The cosmic string generating the acceleration is no longer present, however, the static black hole localized on the brane is nonetheless in an accelerating frame. Further, the brane also intersects the bulk acceleration horizon, which is then interpreted as the dS$_{3}$ cosmological horizon.\footnote{Note that the bulk acceleration horizon is of infinite extent. This can be seen, for example, by setting $\mu=q=0$ and performing a judicious coordinate transformation such that C-metric (\ref{eq:AdS4Ccoord}) becomes Rindler-AdS$_{4}$ \cite{Emparan:2022ijy}. Nonetheless, the positive tension brane induces a compact cosmological horizon.}


The geometry induced on the brane at $x=0$ will result in a metric in $(t,r,\phi)$-coordinates. However, the angular $\phi$ coordinate is not $2\pi$-periodic due to the identification (\ref{eq:conedefidsds}) to remove the bulk conical singularity at $x=x_{1}$.  The metric at $x=0$ may be put into canonically normalized coordinates via the rescaling $(t,r,\phi)\to(\bar{t},\bar{r},\bar{\phi})$, where $t=\Delta\bar{t}$, $r=\Delta^{-1}\bar{r}$, and $\phi=\Delta\bar{\phi}$, where $\Delta$ is reported in (\ref{eq:conedefidsds}). Consequently, $\bar{\phi}$ is periodic in $2\pi$, and the resulting geometry is
\beq \label{eq:regular} 
ds^{2}|_{x=0}=-H(\bar{r})d\bar{t}^{2}+H^{-1}(\bar{r})d\bar{r}^{2}+\bar{r}^{2}d\bar{\phi}^{2}\;,\quad H(\bar{r})=\Delta^{2}-\frac{\bar{r}^{2}}{R_{3}^{2}}-\frac{\Delta^{3}\mu \ell}{\bar{r}}+\frac{q^{2}\ell^{2}}{\bar{r}^{2}}\Delta^{4}\;.\eeq
We note this solution is cosmetically similar to the four-dimensional Reissner-Nordstr\"om de Sitter black hole, which will result in a similar horizon structure. Here, however, the $1/\bar{r}$ and $1/\bar{r}^{2}$ contributions arise due to quantum corrections, as we elucidate momentarily.

The Maxwell field strength also has junction conditions to obey. In particular, let $n^{\mu}$ denote the normal to the brane at $x=0$ (pointing towards increasing values of $x$) and let $e^{\mu}_{i}$ be a basis for tangent vectors to the brane. Then, projecting $F_{ab}$ onto the brane such that $F_{ij}\equiv F_{\mu\nu}e^{\mu}_{i}e^{\nu}_{j}$ and $f_{i}\equiv F_{\mu\nu}e^{\mu}_{i}n^{\nu}$, one has the following junction conditions \cite{Lemos:2021jtm} for a purely tensional brane
\beq 
\begin{split}
&[F_{ij}]=F^{+}_{ij}-F^{-}_{ij}=0\;,\\
&[f_{i}]=f^{+}_{i}-f^{-}_{i}=4\pi j_{i}\;,
\end{split}
\label{eq:junccondEM}\eeq
where $j_{i}$ is the electromagnetic surface current. In coordinates $(\bar{t},\bar{r},\bar{\phi})$, the projected components of the electromagnetic tensor are
\beq F_{\bar{r}\bar{t}}=\frac{2q\ell}{\ell_{\star}\bar{r}^{2}}\Delta^{2}\;,\quad f_{\bar{\phi}}=0\;.\label{eq:projcomps}\eeq
Using the junction conditions (\ref{eq:junccondEM}), the induced current density vanishes, $j^{\bar{\phi}}=0$.\footnote{Had we included magnetic charge, one would find $f_{\bar{\phi}}=-\frac{2g\ell}{\ell_{\star}\bar{r}}\Delta^{2}$ such that $j^{\bar{\phi}}=\frac{g\ell\Delta^{2}}{\pi\ell_{\star}\bar{r}^{3}}$ \cite{Climent:2024nuj}.}


\subsection{Quantum black hole interpretation}

\subsection*{Induced brane theory}

Thus far we have only described an exact black hole solution to Einstein-Maxwell-AdS$_{4}$ gravity coupled to an end-of-the-world Randall-Sundrum brane. That is, we have taken a bulk perspective, 
\beq I=I_{\text{bulk}}+I_{\text{bdry}}+I_{\text{brane}}\;,\label{eq:totalactionbulk}\eeq
where $I_{\text{bulk}}$ is the Einstein-Maxwell action (\ref{eq:bulk_action}) and $I_{\text{brane}}$ is the tensional brane action (\ref{eq:braneact}). Moreover, the action $I_{\text{bdry}}$ consists of the standard Gibbons-Hawking-York (GHY) term necessary for the well-posedness of the variational problem of Einstein gravity with a Dirichlet boundary, and an analogous boundary term to make the Dirichlet variational problem of the Maxwell term well-posed (see below). 

There is also a brane perspective, characterized by an effective induced theory of gravity. As we outlined in the introduction, the specific theory on the brane follows from integrating out the bulk from the AdS$_{4}$ boundary up to the ETW brane, analogous to holographic renormalization \cite{Kraus:1999di,Emparan:1999pm,deHaro:2000vlm,Skenderis:2002wp,Papadimitriou:2004ap}. Skipping the details, the action (\ref{eq:totalactionbulk}) becomes 
\beq I=\frac{\ell_{4}}{8\pi G_{4}}\int_{\mathcal{B}}d^{3}x\sqrt{-h}\left[\frac{4}{\ell_{4}}\left(\frac{1}{\ell}-\frac{1}{\ell_{4}}\right)+R+\ell_{4}^{2}\left(\frac{3}{8}R^{2}-R_{ij}^{2}\right)+...\right]+I_{\text{EM}}+I_{\text{CFT}}\;.\label{eq:inducactqbtzchar}\eeq
The first term is a purely gravitational action unaffected by the bulk Maxwell terms, where the ellipsis corresponds to an tower of higher-derivative terms\footnote{See, e.g., \cite{Bueno:2022log} for specific expressions of the higher-derivative action and \cite{Bueno:2023qpr} for additional analysis of its non-local and massive behavior.} and $I_{\text{CFT}}$ is a large-$c$ holographic CFT$_{3}$ with a UV cutoff (due to the presence of the brane). Combined, these two terms characterize the induced brane theory for neutral quantum black holes \cite{Emparan:2020znc,Emparan:2022ijy,Panella:2023lsi}. Including the bulk Maxwell term enriches the induced theory with the electromagnetic term $I_{\text{EM}}$ is 
\beq I_{\text{EM}}=2\int_{\mathcal{B}} d^{3}x\sqrt{-h}A_{i}j^{i}+I^{\text{ct}}_{\text{EM}}\;.\eeq
The first contribution is the boundary term necessary to keep the bulk Maxwell Dirichlet variational problem well-posed, which in our case is zero because we assume there is no magnetic charge. The second contribution describe local counterterms associated with the four-dimensional bulk Maxwell action that are included in holographic renormalization\footnote{In addition to the local counterterms in pure gravity, $p$-form fields $F_{p}$ (where $p=2$ corresponds to Maxwell) may require local counterterm subtraction. As reported in \cite{Taylor:2000xw}, for a $d+1$-dimensional bulk, when $d<2p$ there are no divergences while a logarithmic divergence appears for $d=2p$ and there will be divergences for $d>2p$. Further, for $d=2p+2n$ with $n\in\mathbb{Z}^{+}$, derivatives of $F_{p}$ and its coupling to curvature appear in the conformal anomaly such that counterterms are needed for $d>2p+2n$. Thus, the four-dimensional Maxwell action has no divergences as the IR cutoff $\epsilon\to0$. Such terms, however, contribute on the brane because the brane effective action keeps the cutoff finite and non-zero.} \cite{Taylor:2000xw}
\beq
\begin{split}
 I^{\text{ct}}_{\text{EM}}=\frac{\ell_{4}\ell^{2}_{\ast}}{8\pi G_{4}}&\int d^{3}x\sqrt{-h}\biggr[-\frac{5}{16}F^{2}+\ell_{4}^{2}\biggr(\frac{1}{288}RF^{2}-\frac{5}{8}R^{i}_{\;j}F_{ik}F^{jk}\\
&+\frac{3}{98}F^{ij}(\nabla_{j}\nabla^{k}F_{ki}-\nabla_{i}\nabla^{k}F_{kj})+\frac{5}{24}\nabla_{i}F^{ij}\nabla_{k}F^{k}_{\;j}\biggr)+\mathcal{O}(\ell_{4}^{3})\biggr]\;.
\end{split}
\eeq
We see the higher-derivative contributions are multiplied by increasing powers of the bulk AdS length $\ell_{4}$. 

Altogether, the induced theory (\ref{eq:inducactqbtzchar}) may thus be interpreted as a three-dimensional semi-classical theory of gravity, with an effective three-dimensional Newton's constant $G_{3}$, cosmological constant $\Lambda_{3}$, and gauge coupling $\tilde{\ell}_{\star}$ (or $g_{3}$), 
\beq 
\begin{split}
&G_{3}\equiv\frac{G_{4}}{2\ell_{4}}\;,\\
&\frac{2}{L_{3}^{2}}\equiv \frac4{\ell_4}\left(\frac1{\ell}-\frac1{\ell_{4}}\right)\;,\\
&\tilde{\ell}_{\star}^{2}=\frac{16\pi G_{3}}{g_{3}^{2}}\;,\quad g_{3}^{2}=\frac{2}{5}\frac{g_{\ast}^{2}}{\ell_{4}}\;,\\
\end{split}
\label{eq:branecc}\eeq
such that $\ell_{\star}^{2}=4\tilde{\ell}^{2}_{\ast}/5$. The effective three-dimensional theory will be valid when $\ell_{4}\ll L_{3}$.
Via the relation of the bulk length scale (\ref{eq:L4bulk}), this implies $\ell\sim\ell_{4}\ll R_{3}$. This amounts to placing the cut-off brane near the (now removed) portion of the AdS$_{4}$ boundary. Since generally the bulk length $\ell_{4}$ plays the role of a UV cutoff for the holographic CFT$_{3}$, the parameter $\ell$ may be treated as a small UV cutoff length scale. Regarding $\ell/R_{3}$ as a small expansion parameter of the effective theory, the induced cosmological constant (\ref{eq:branecc}) satisfies $L_{3}\approx R_{3}$, though technically the physical curvature radius of the brane is not the same as $L_{3}$ due to the higher-derivative contributions. Moreover, notice that in the limit $\ell\to0$, the coupling (\ref{eq:branecc}) $g_{3}\to\infty$ becomes non-dynamical. We will see momentarily how this charge contribution to the metric disappears in the same limit.

Then, treating $\ell\ll R_{3}$, the effective theory on the brane (\ref{eq:inducactqbtzchar}) becomes \cite{Climent:2024nuj}
\beq
\begin{split}
 I&=\frac{1}{16\pi G_{3}}\int_{\mathcal{B}} d^{3}x\sqrt{-h}\biggr[R-\frac{2}{R_{3}^{2}}-\frac{\tilde{\ell}_{\star}^{2}}{4}F^{2}+16\pi G_{3}A_{i}j^{i}\\
&+\ell^{2}\left(\frac{3}{8}R^{2}-R_{ij}^{2}\right)+\frac{4}{5}\ell^{2}\tilde{\ell}^{2}_{\ast}\biggr(\frac{1}{288}RF^{2}-\frac{5}{8}R^{i}_{\;j}F_{ik}F^{jk}\\
&+\frac{3}{98}F^{ij}(\nabla_{j}\nabla^{k}F_{ki}-\nabla_{i}\nabla^{k}F_{kj})+\frac{5}{24}\nabla_{i}F^{ij}\nabla_{k}F^{k}_{\;j}\biggr)+\mathcal{O}(\ell^{3})\biggr]+I_{\text{CFT}}\;.
\end{split}
\label{eq:indactbranecharge}\eeq
With $\ell$ being small, the higher-derivative terms may be viewed as corrections to the leading Einstein-Hilbert action. The purely higher-curvature gravity contributions arise at order $\mathcal{O}(\ell^{2})$ and (though not immediately apparent) the non-minimally coupled terms enter at order $\mathcal{O}(\ell^{3})$. Meanwhile if we normalize the central charge $c$ of the boundary CFT to be $c=\ell_{4}^{2}/G_{4}$, as is standard, then in this small $\ell$ approximation it follows 
\beq 2cG_{3}\approx \ell\;.\eeq
Thus, the effects of the CFT enter at order $\mathcal{O}(\ell)$. Further, notice, for fixed $c$, gravity on the brane becomes weak ($G_{3}\to0$) as $\ell\to0$, such that there is no backreaction due to the CFT.  In this way, the parameter $\ell/R_{3}$ characterizes the strength of backreaction, with $\ell/R_{3}\ll1$ indicating small backreaction -- the regime we are primarily interested in. Altogether, the induced theory (\ref{eq:indactbranecharge}) may thus interpreted as a semi-classical theory of gravity  where the higher-derivative corrections incorporate backreaction effects due to the CFT on the brane.

The metric equations of motion of the induced theory to order $\mathcal{O}(\ell^{2})$ are as the neutral case \cite{Emparan:2022ijy,Panella:2023lsi}, except now with an additional contribution coming from the $F^{2}$ contribution in the action. In particular, 
\beq 
\begin{split}
8\pi G_{3}\langle T_{ij}\rangle&=G_{ij}+\frac{1}{L_{3}^{2}}h_{ij}-\frac{\tilde{\ell}^{2}_{\ast}}{2}\left(F^{k}_{i}F_{jk}-\frac{1}{4}h_{ij}F^{2}\right)+16\pi G_{3} A_{k}j^{k}h_{ij}+...\;,
\end{split}
\label{eq:stresstenscqbtzlead}\eeq
where $G_{ij}$ is the three-dimensional Einstein tensor and the ellipsis refers to terms at higher order in $\ell$. Further, varying with respect to the gauge field $A_{i}$, we find the analog of the semi-classical Maxwell equations, 
\beq
\begin{split}
\langle J^{j}\rangle&=j^{j}+\frac{\tilde{\ell}^{2}_{\ast}}{16\pi G_{3}}\biggr\{\nabla_{i}F^{ji}+\frac{16}{5}\ell^{2}\biggr(-\frac{1}{72}R\nabla_{i}F^{ji}+\frac{11}{18}F^{j}_{\;\,i}\nabla^{i}R+\frac{209}{294}R^{ij}\nabla_{k}F_{i}^{\;k}\\
&+\frac{5}{4}R^{ik}\nabla_{k}F^{j}_{\;\,i}+\frac{5}{4}F^{ik}\nabla_{k}R^{j}_{\;i}+\frac{317}{588}\nabla_{i}\nabla^{i}\nabla_{k}F^{jk}+\frac{317}{588}\nabla^{j}\nabla^{k}\nabla_{i}F^{ik}\biggr)+\mathcal{O}(\ell^{3})\biggr\}\;.
\end{split}
\label{eq:sccurrentdens}\eeq
Solutions to this complicated theory may be understood as quantum-corrected geometries \cite{Emparan:2002px}. Luckily, we can use the correspondence between the bulk and brane pictures and find exact solutions without explicitly solving the induced equations of motion. One such solution is the family of charged quantum BTZ (qBTZ) black holes \cite{Climent:2024nuj}. Below we will interpret the  brane geometry (\ref{eq:regular}) as a charged quantum dS$_{3}$ black hole.

\subsection*{Quantum black hole on the brane}

Let us return to the geometry (\ref{eq:regular}) on the brane at $x=0$. We may identify the black hole mass $M$ via
\beq 8\mathcal{G}_{3}M\equiv 1-\Delta^{2}=1-\frac{4x_{1}^{2}}{(3-x_{1}^{2}+\gamma^{2})^{2}}\;,\label{eq:massdef}\eeq
where $\gamma\equiv qx_{1}^{2}$. Here $\mathcal{G}_{3}\equiv \ell_{4} G_{3}/\ell=G_{3}/\sqrt{1-(\ell/R_{3})^{2}}$ denotes a ``renormalized'' Newton's constant accounting for modifications to the mass due to the presence of higher-derivative corrections appearing in the induced action \cite{Cremonini:2009ih}. At leading order in small backreaction, $\mathcal{G}_{3}\approx G_{3}$. We further define form functions $F(M,q)$ and $Z(M,q)$ as
\beq F(M,q)\equiv\mu\Delta^{3}=\frac{8(1-x_{1}^{2}-\gamma^{2})}{(3-x_{1}^{2}+\gamma^{2})^{3}}\;,\label{eq:FMdef}\eeq
\beq Z(M,q)\equiv q^{2}\Delta^{4}=\frac{16\gamma^{2}}{(3-x_{1}^{2}+\gamma^{2})^{4}}\;.\label{eq:Q2def}\eeq
With these identifications the brane metric (\ref{eq:regular}) becomes
\beq ds^{2}=-H(\bar{r})d\bar{t}^{2}+H^{-1}(\bar{r})d\bar{r}^{2}+\bar{r}^{2}d\bar{\phi}^{2}\;,\quad H(\bar{r})=1-8\mathcal{G}_{3}M-\frac{\bar{r}^{2}}{R_{3}^{2}}-\frac{\ell F}{\bar{r}}+\frac{\ell^{2}Z}{\bar{r}^{2}}\;.\label{eq:qsdsc}\eeq
Since the brane black hole is an exact solution to the classical bulk gravity theory, we are guaranteed the geometry (\ref{eq:qsdsc}) is an exact solution to the holographically induced theory of gravity, including the entire tower of higher-derivative corrections to the Einstein-Hilbert term. We can thus think of the brane metric (\ref{eq:qsdsc}) as a \emph{quantum} black hole, i.e., a quantum-corrected geometry accounting for all orders of semi-classical backreaction due to the CFT$_{3}$. 

It is worth highlighting that since the three-dimensional Planck length is $L_{\text{P}}=G_3$ (with $\hbar=1$), then scale $\ell$ is approximately
\beq \ell\approx  2cL_{\text{P}}\;,\eeq
with corrections suppressed by $\mathcal{O}(cL_{\text{P}}/R_{3})^{2}$. Thence, since our holographic construction demands we have a large number of CFT degrees of freedom, $c\gg1$, the length scale $\ell$ is much larger than the Planck scale, $\ell \gg L_{\text{P}}$. Consequently, the constructed quantum black holes are of size $\propto cL_{\text{P}}$, and we may consistently ignore quantum gravity effects. 

Substituting the solution (\ref{eq:qsdsc}) into the quantum stress-tensor (\ref{eq:stresstenscqbtzlead}) yields, 
\beq \langle T^{i}_{\;j}\rangle\approx \frac{c}{8\pi}\frac{F(M)}{\bar{r}^{3}}\text{diag}(1,1,-2)+\mathcal{O}(\ell^{2})+...\;,\label{eq:stresstensbraneq0}\eeq
where the effects of $q$ in $F(M,q)$ enter at quadratic order in $\ell$, and thus at leading order $F(M,q)\approx F(M)$, with $F(M)$ being the same form function as in the quantum Schwarzschild-de Sitter solution (Eq. (4.14) of \cite{Emparan:2022ijy}). Further, note the structural similarity between the stress-tensor (\ref{eq:stresstensbraneq0}) and that of the defect (\ref{eq:expvalTmunudefect}). Notably, at fixed central charge $c$, the two stress-tensors coincide in the limit of vanishing backreaction; this confirms the double-Wick rotated geometries (\ref{eq:bdry_defect}) allow one to compute the stress-tensor of a chemical defect which then backreacts on the geometry.

In principle we could compute higher-order contributions to $\langle T^{i}_{\;j}\rangle$, however, it is tedious and not particularly enlightening, except that the stress-tensor is not traceless at $\mathcal{O}(\ell^{2})$, a consequence of the UV cutoff breaking conformal invariance.  Moreover, using the projected components of the Maxwell field strength (\ref{eq:projcomps}), the leading order $\mathcal{O}(\ell)$ contribution to the semi-classical current density (\ref{eq:sccurrentdens}) has components
\beq
\begin{split}
&\langle J^{\bar{t}}\rangle=-\frac{\ell\tilde{\ell}_{\star}^{2}}{8\pi G_{3}\ell_{\star}}\frac{q\Delta^{2}}{\bar{r}^{3}}\propto \frac{q\ell\sqrt{c}}{g_{\star}\bar{r}^{3}}\;,\quad \langle J^{\bar{\phi}}\rangle=0\;.
\end{split}
\eeq
Interestingly, the temporal component vanishes in the limit $\ell\to0$.\footnote{Including magnetic charge, one finds $\langle J^{\bar{\phi}}\rangle=\frac{\ell}{\ell_{\star}}\frac{g\Delta^{2}}{\pi\bar{r}^{3}}\propto \frac{gg_{\star}\sqrt{c}}{\bar{r}^{3}}$, which is independent of the backreaction parameter $\ell$ \cite{Climent:2024nuj}.} This is consistent with the three-dimensional charged defects in conical dS$_{3}$ we described in Section~\ref{app:ds3_defects}. 

Notice that, unlike the rotating case \cite{Panella:2023lsi}, the limit of vanishing backreaction does not return a classically dS$_{3}$ geometry electrically charged under three-dimensional Maxwell theory. Indeed, in three-dimensions, the gauge field $A_{t}$ has a logarithmic dependence, $A_{t}\sim q\log(r)$, which produces a logarithmic correction to the three-dimensional blackening factor (as in the case of the charged BTZ black hole \cite{Martinez:1999qi}). As pointed out in \cite{Climent:2024nuj}, the lack of logarithmic behavior arises from the fact the bulk four-dimensional gauge field does not localize on the brane in the same way as a gravity. Nonetheless, the quantum black hole (\ref{eq:qsdsc}) is charged. From the brane perspective, to compute the charge $Q$, one must perform a resummation of the whole infinite tower of higher-derivative terms appearing in the induced theory (\ref{eq:inducactqbtzchar}). Fortunately, the bulk theory performs this resummation, and the charge of the brane black hole is identified with the electric charge of the bulk black hole \cite{Climent:2024nuj} (see also \cite{Emparan:2000fn})
\beq Q=\frac{2}{g_{\star}^{2}}\int\star F=\frac{2q\ell}{g_{\star}^{2}\ell_{\star}}\int_{0}^{2\pi\Delta}\hspace{-2mm}d\phi\int_{0}^{x_{1}}\hspace{-1mm}dx=\frac{8\pi\ell q \Delta x_{1}}{g_{\star}^{2}\ell_{\star}}\;,\label{eq:chargebulk}\eeq
where the factor of two in the first equality is due to the $\mathbb{Z}_{2}$ symmetry and $\star F=r^{2}F_{rt}d\phi dx$ refers to the Hodge dual of the bulk Maxwell tensor. We will see further evidence this is the correct form of the brane black hole charge when we study the horizon thermodynamics. Notice that, contrary to what happens with higher-dimensional charged black holes, it is not $Q^{2}$ which appears in the $1/\bar{r}^{2}$ term in the blackening factor (\ref{eq:qsdsc}). 

Associated with the electric charge is its electrostatic potential $\Phi$. Since the bulk Einstein-Maxwell theory has a dual interpretation in terms of a CFT$_{3}$ with a chemical potential, it is natural to refer to $\Phi$ as a chemical potential. Specifically, it is equal to the boundary value of the bulk gauge field (\ref{eq:gauge_Cmetric}) restricted to the brane
\begin{equation} \label{eq:chemicalpotentialbulk}
   \Phi_{i} \equiv \lim_{r\to\infty}A_{b}\bar{\zeta}^{b}=\lim_{r\to\infty}A_{\bar{t}} = \dfrac{2\ell q \Delta}{r_{i}\ell_\star} \, ,
\end{equation}
Here $\bar{\zeta}^b = \partial^b_{\bar{t}}$, is the timelike Killing vector generating each horizon; the index on $\Phi_{i}$ and $r_{i}$ refers to either the black hole or cosmological horizon radii, $r_{h}$ or $r_{c}$, as we will treat the outer black hole and cosmological horizons separately. Notice that in the limit of vanishing backreaction $\ell\to0$, the charge and chemical potential vanish. This is indicative of the fact the charge of the brane black hole is a quantum effect, and the charged quantum black hole does not reduce to classical electrically charged dS$_{3}$.


\subsection{Extremal, Nariai, and ultracold black holes}
Let us now describe the horizon structure of the quantum black hole (\ref{eq:qsdsc}). The roots of $H(\bar{r})$ of the three-dimensional geometry (\ref{eq:regular}) characterize the Killing horizons of the black hole solution, generated by the time-translation Killing vector field $\partial_{\bar{t}}$. Defining the quartic polynomial $Q(\bar{r})\equiv \bar{r}^2 H(\bar{r})$, there are in general four, two, or zero real roots. Focusing on the case of four real roots, three are positive, $\bar{r}_{c}\geq \bar{r}_{+}\geq \bar{r}_{-}$, corresponding to the cosmological, outer, and inner black hole horizons, respectively. The fourth root, $\bar{r}_{n}=-(\bar{r}_{c}+\bar{r}_{+}+\bar{r}_{-})<0$ is the unphysical negative horizon lying behind the curvature singularity at $\bar{r}=0$. Using $H(\bar{r}_{c})=H(\bar{r}_{\pm})=0$, it is straightforward to express
\beq
\begin{split}
&R_{3}^{2}\Delta^{2}=\bar{r}_{c}^{2}+\bar{r}_{+}^{2}+\bar{r}_{c}\bar{r}_{+}+\bar{r}_{-}(\bar{r}_{c}+\bar{r}_{+}+\bar{r}_{-})\;,\\
&\mu\ell \Delta =\frac{(\bar{r}_{c}+\bar{r}_{+})(\bar{r}_{c}+\bar{r}_{-})(\bar{r}_{+}+\bar{r}_{-})}{R_{3}^{2}\Delta^{2}}\;,\\
&q^{2}\ell^{2}\Delta^{2}=\frac{\bar{r}_{c}\bar{r}_{+}\bar{r}_{-}(\bar{r}_{c}+\bar{r}_{+}+\bar{r}_{-})}{R_{3}^{2}\Delta^{2}}\;.
\end{split}
\label{eq:paramsintermsofri}\eeq

Each horizon
has an associated surface gravity $\kappa$, defined by $\bar{\zeta}^b \nabla_c\bar{\zeta}^c = \kappa \bar{\zeta}^c$. Using the convenient factorization of the blackening factor,
\beq
 H(\bar{r}) = \frac{1}{\bar{r}^2 R_3^2} (\bar{r}_c - \bar{r}) (\bar{r} - \bar{r}_+) (\bar{r} - \bar{r}_-) (\bar{r} + \bar{r}_+ + \bar{r}_- + \bar{r}_c ) \ ,
\eeq
the surface gravities at the three horizons are
\beq
\begin{split}
&\kappa_{c}=-\frac{1}{2\bar{r}_c^2 R_3^2}(\bar{r}_{c}-\bar{r}_{+})(\bar{r}_{c}-\bar{r}_{-})(\bar{r}_{+}+\bar{r}_{-}+2\bar{r}_{c})\;,\\
&\kappa_{+}=\frac{1}{2\bar{r}_+^2R_{3}^{2}}(\bar{r}_{c}-\bar{r}_{+})(\bar{r}_{+}-\bar{r}_{-})(\bar{r}_{c}+\bar{r}_{-}+2\bar{r}_{+})\;,\\
&\kappa_{-}=-\frac{1}{2\bar{r}_-^2 R_{3}^{2}}(\bar{r}_{c}-\bar{r}_{-})(\bar{r}_{+}-\bar{r}_{-})(\bar{r}_{c}+\bar{r}_{+}+2\bar{r}_{-})\;.
\end{split}
\label{eq:surfgravs}\eeq
where we used $2\kappa_{i}=|H'(\bar{r}_{i})|$.

There are three cases when geometry has degenerate horizons: (i) the extremal or ``cold'' limit, where $\bar{r}_{+}=\bar{r}_{-}$; (ii) the Nariai limit, with $\bar{r}_{c}=\bar{r}_{+}$, and (iii) the ``ultracold'' limit, where $\bar{r}_{c}=\bar{r}_{+}=\bar{r}_{-}$, as we now describe.


\subsubsection*{Extremal black hole: $\bar{r}_{+}=\bar{r}_{-}$}
The extremal, or cold \cite{Romans:1991nq}, black hole corresponds to the limit in which the outer and inner black hole horizons coincide. The name comes from the fact that, in this regime, the surface gravity $\kappa_+$ (and, consequently, the temperature $T_+$) associated with the outer horizon goes to zero.
In this limit,  the key black hole parameters can be rewritten as\footnote{These follow from substituting $r_{-}=r_{+}$ into parameters (\ref{eq:paramsintermsofri}) and rearranging.}
\beq
\label{eq:extremal}
 q^2\ell^2 \Delta^4 =\frac{\bar{r}_+^2}{R_3^2} (R_3^2 \Delta^2- 3\bar{r}_+^2) \ , \qquad \mu \ell \Delta^3 = \frac{2 \bar{r}_+}{R_3^2} (R_3^2 \Delta^2-2 \bar{r}_+^2) \ .
\eeq
Note that the black hole interior becomes physically inaccessible from the rest of the spacetime, since the horizon is an infinite proper distance away from all points in the exterior. 

The near-horizon geometry of the near-extremal quantum black hole has a similar structure of the classical four-dimensional RN-de Sitter black hole. To see this, perform the change of coordinates
 (following \cite{Bardeen:1999px} and \cite{Hartman:2008pb})
\beq
\bar{r} = \bar{r}_+ + \lambda \sqrt{\Gamma}\rho \ ,  \qquad \bar{t} = \frac{\sqrt{\Gamma}\tau}{\lambda} \ ,
\eeq
with $\lambda>0$ and 
\beq
\Gamma = \frac{\bar{r}_+^2 R_3^2}{R_3^2 \Delta^2-6\bar{r}_+^2}\;.
\eeq
Carefully sending $\lambda \to 0$ we obtain
\beq \label{eq:extAdS2}
ds^2_{\text{ex}}= \Gamma \left(-\rho^2 d \tau^2 + \frac{d\rho^2}{\rho^2} \right) + \bar{r}_+^2 d \bar{\phi}^2 \,,
\eeq
having the form of the product manifold $\text{AdS}_{2}\times S^{1}$, with isometry group SL$(2,\mathbb{R})\times U(1)$. It would be particularly interesting to work out the low-energy description of the near-extremal solutions in the near-horizon regime. These solutions are likely captured by a deformation of AdS$_{2}$ Jackiw-Teitelboim (JT) gravity. From the brane perspective, deformations will arise due to the higher-curvature terms present in the brane action. From the four-dimensional bulk perspective, these deformations can be understood to arise because the extremal limit of the C-metric is distorted due to acceleration. That is, the extremal geometry of the C-metric is still topologically AdS$_2 \times S^2$, however, the $S^2$ is not homogeneous~\cite{Dias:2003up, Kolanowski:2023hvh, Climent:2024nuj}.



\subsubsection*{Charged Nariai black hole: $\bar{r}_c = \bar{r}_+$}

The Nariai black hole is the largest black hole which can fit inside the cosmological horizon, with horizon radius $\bar{r}_c = \bar{r}_+ \equiv \bar{r}_N$, without introducing naked singularities. Via the parameter relations (\ref{eq:paramsintermsofri}), here
\beq
\label{eq:nairiai}
q^2 \ell^2 \Delta^4 = \frac{\bar{r}_N^2}{R_3^2}(R_3^2 \Delta^2-3\bar{r}_N^2) \ , \qquad \mu \ell  \Delta^{3}=\frac{2\bar{r}_N}{R_3^2}(R_3^2 \Delta^2-2 \bar{r}_N^2) .
\eeq
When the charge vanishes ($q=0$), then $\bar{r}_N=R_3 \Delta/\sqrt{3}$, the Nariai limit for a (quantum) Schwarzschild-de Sitter black hole \cite{Emparan:2022ijy}. 

To probe the near-horizon geometry, perform the change of coordinates
\beq
\bar{r} = \bar{r}_N + \epsilon \rho \ , \qquad \bar{t} = \frac{\Gamma \tau}{\epsilon} \ 
\eeq
for $\epsilon>0$ and
\beq
\Gamma = \frac{R_3^2 \bar{r}_N^2}{6\bar{r}_N^2-R_3^2 \Delta^2} \ .
\eeq
Sending $\epsilon \to 0$ yields
\beq
ds^2_N = \Gamma \left(-(1-\rho^2)d\tau^2 + \frac{d\rho^2}{(1-\rho^2)}\right)+r_N^2 d\bar{\phi}^2 \ .
\label{eq:Nariaigeom}\eeq
 Thus, the (near-horizon) Nariai geometry takes the form $\text{dS}_{2}\times S^{1}$ (here expressed in static patch coordinates), analogous to the four-dimensional charged Nariai solution.
 The isometry group remains SL$(2,\mathbb{R})\times U(1)$. The static patch observer is restricted to the region $\rho \in (-1,1)$, where $\rho_{+}=-1$ and $\rho_{c}=1$ are, respectively, the locations of the black hole and cosmological horizon. The low-energy effective dynamics is likely characterized by a deformation of dS$_{2}$ JT gravity.

\subsubsection*{Ultracold black hole: $\bar{r}_c=\bar{r}_+=\bar{r}_-$}
The limit in which all the three horizons coincide ($\bar{r}_c=\bar{r}_+=\bar{r}_-\equiv \bar{r}_{\text{u}}$) is known as the ultracold limit. Its near-horizon geometry can uncovered as follows. Start with the Nariai metric (\ref{eq:Nariaigeom}). The limit
$\bar{r}_N = \bar{r}_-$ is singular. Thus, rescale coordinates $(\tau,\rho)$ 
\beq
\rho = \sqrt{\frac{2\bar{r}_{\text{u}}-\delta}{R_3}}X \ , \qquad \tau = \sqrt{\frac{R_3}{2\bar{r}_{\text{u}}-\delta}}\frac{R_3 \bar{r}_{\text{u}}}{4}T 
\eeq
and subsequently take the limit $\delta \to 2r_{\text{u}}$. The resulting geometry is then
\beq
ds^2_{\text{u}} = \frac{R_3 \bar{r}_{\text{u}}}{4}(-dT^2+dX^2) + \bar{r}_{\text{u}}^2 d\bar{\phi}^2 \,,
\eeq
the product manifold $\text{Mink}_{2}\times S^{1}$. In the limit of vanishing charge, there is no ultracold solution and the geometry returns to neutral Nariai. The low-energy description is likely captured by a deformation of flat JT gravity.



\subsection*{The shark fin phase diagram}

It is useful to pause at this point and study the landscape of solutions, following the analysis in \cite{Morvan:2022aon}. On physical grounds, we would like to avoid naked singularities.  These corresponds to points in parameter space where the non-negative roots $\bar{r}_-$, $\bar{r}_+$ and $\bar{r}_c$ become complex. That is, the discriminant of the quartic equation $H(\bar{r})=0$ becomes negative. Let us first look at the naive metric. It is easy to see the domain of existence, as for the four-dimensional Reissner-Nordstr\"om de Sitter black hole, is then
\begin{equation}
    \mu_\text{E} \leq \mu \leq \mu_\text{N} \ , \qquad |q_\text{N}| \leq |q| \leq |q_\text{E}| \ ,
\end{equation}
where $\mu_{\text{E}},q_{\text{E}}$ denote the values of the parameters $\mu$ and $q$ of the extremal black hole, and similarly for $\mu_{\text{N}}, q_{\text{N}}$. 

From the perspective of a brane observer, we are simply restating that the only allowed solutions for a fixed $q$ are those bounded on the left by the extremal black hole and on the right by the Nariai solution (as a function of $\mu$). The non-trivial difference with respect to the analysis of the classical RN black holes is, however, that the metric parameters are non-trivially related to the physical mass and charge by two distinct functions of $x_1$. As detailed in Appendix \ref{app:range_bulk}, we find for fixed $q$, the parameter $x_1$ is bounded below by its value at the Nariai solution $x_1^{\text{N}}$. Since $\Delta$ is monotonic in $x_1$ in the physical parameter range, the physical masses are bounded above by the Nariai mass $M_{\text{N}}$. We can repeat the argument for the extremal solution, showing that the edges of the space of allowed solutions, in terms of the physical variables, are
\begin{equation}
    M_\text{E} \leq M \leq M_\text{N} \ , \qquad |Q_\text{N}| \leq |Q| \leq |Q_\text{E}| \ ,
\end{equation}
with, respectively, extremal and Nariai mass, $M_{\text{E}}$ and $M_{\text{N}}$, and similarly for charge.

\begin{figure}[t!]
  \centering
  \begin{minipage}[t]{0.47\textwidth}
    \includegraphics[width=\textwidth]{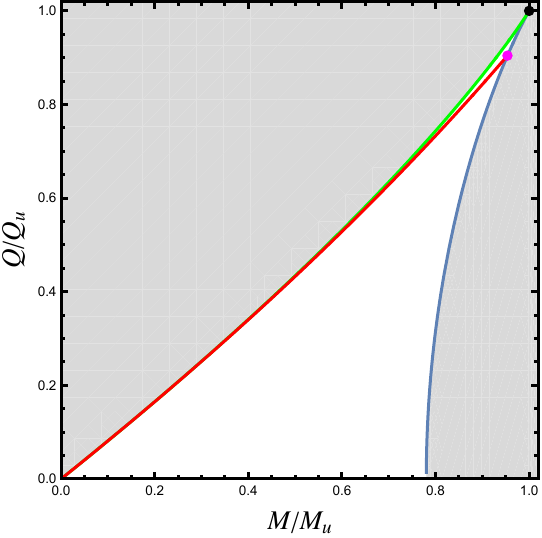}
  \end{minipage}
  \hfill
  \begin{minipage}[t]{0.47\textwidth}
    \includegraphics[width=\textwidth]{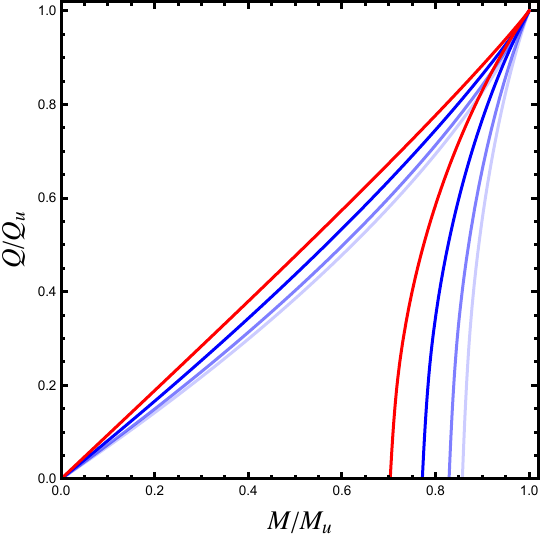}
  \end{minipage}
  \caption{\small \textbf{Left:} Phase diagram of quantum black holes parametrized by charge $Q$ and mass $M$, normalized with respect to their ultracold values for $\nu =  1/2$. Blue and green curves correspond to the Nariai and extremal families. The the red line corresponds to the lukewarm black holes, namely, where outer black hole and cosmological horizon temperatures are equal (cf. Fig. \ref{fig:T}). States in the gray region have naked singularities. The black and magenta dots represent the ultracold and Nariai-lukewarm solutions. \textbf{Right:} The sharkfin diagram for different values of the backreaction parameter $\nu$. From dark to light we have $\nu = 0.99,1/3,1/5$, whilst the red curve represents the sharkfin for the classical (3+1)-dimensional RN dS black hole.}
  \label{fig:sharkfin}
\end{figure}

Visually, a plot of charge versus mass traces the characteristic ``shark fin'' diagram of classical charged de Sitter black holes (Figure \ref{fig:sharkfin}). For readability we have normalized both the physical mass and the charge with respect to the ultracold solution, $Q_{\text{u}}, M_{\text{u}}$. The edges of the sharkfin are obtained by plotting $M(q_{\text{E}/\text{N}},\mu_{\text{E}/\text{N}})$ and $Q(q_{\text{E}/\text{N}},\mu_{\text{E}/\text{N}})$ as parametric curves using \eqref{eq:nairiai} and \eqref{eq:extremal}, where the outer horizon radius is used as a parameter. Recall that its range of values in the extremal limit is $r_{\text{E}} \in [0,r_{\text{u}}]$, whilst for the Nariai limit $r_\text{N} \in [r_{\text{u}},r_\text{N}|_{q=0}]$. Changing the value of $\nu$ smoothly deforms the edges of the diagram by mapping the metric parameters $q$ and $\mu$ to different physical mass and charge. As $\nu \to \infty$ the solution becomes more and more 4-dimensional, with the phase diagram converging to the one of the classical 4D RN black hole.

\section{Horizon thermodynamics}\label{sec:thermo}

As the bulk geometry imprints itself on the brane, so too does the bulk horizon thermodynamics. To proceed, it behooves us to introduce dimensionless parameters 
\beq z\equiv\frac{R_{3}}{x_{1}r_{i}}\;,\quad \nu\equiv\frac{\ell}{R_{3}}\;, \label{eq:defznu}\eeq
where $r_{i}$ is a positive real root of the blackening factor $H(r)$, representing each of the inner and outer black hole horizons and the cosmological horizon. Rearranging $H(r_{i})=0$, we then express parameters $x_{1}$, $\mu$, and $r_{i}$ in terms of $z,\nu$, and $\gamma\equiv qx_{1}^{2}$,
\beq 
\begin{split}
&x_{1}^{2}=\frac{(1+\nu z^{3}-\gamma^{2}\nu z^{3}-\gamma^{2}\nu^{2}z^{4})}{z^{2}(1+\nu z)}\;,\\
&r_{i}^{2}=\frac{R_{3}^{2}(1+\nu z)}{(1+\nu z^{3}-\gamma^{2}\nu z^{3}-\gamma^{2}\nu^{2}z^{4})}\;,\\
&\mu x_{1}=-\frac{(1-z^{2}+\gamma^{2}z^{2}-\gamma^{2}\nu^{2}z^{4})}{(1+\nu z^{3}-\gamma^{2}\nu z^{3}-\gamma^{2}\nu^{2}z^{4})}\;.
\end{split}
\eeq
Consequently, the mass $M$ (\ref{eq:massdef}), and form functions $F(M,q)$ (\ref{eq:FMdef}), and $Z(M,q)$ (\ref{eq:Q2def}) are
\beq M=\frac{\sqrt{1-\nu^{2}}}{8G_{3}}\left(1-\frac{4z^{2}(1+\nu z)(1+\nu z^{3}[1-\gamma^{2}(1+\nu z)])}{(1-z^{2}[3+2\nu z+\gamma^{2}(1+\nu z)^{2}])^{2}}\right)\;,\label{eq:massz}\eeq
\beq F(M, q)=\frac{8z^{4}(1+\nu z)^{2}(1-z^{2}[1+\gamma^{2}(\nu^{2}z^{2}-1)])}{(1-z^{2}[3+2\nu z+\gamma^{2}(1+\nu z)^{2}])^{3}}\;,\label{eq:FMz}\eeq
\beq Z(M,q)=\frac{16\gamma^{2}z^{8}(1+\nu z)^{4}}{(1-z^{2}[3+2\nu z+\gamma^{2}(1+\nu z)^{2}])^{4}}\;.\label{eq:Qsqz}\eeq
Meanwhile, the charge $Q$ (\ref{eq:chargebulk}) and chemical potential $\Phi_i$ \eqref{eq:chemicalpotentialbulk} are
\beq Q=-\sqrt{\frac{16\pi}{5g_{3}^{2}G_{3}}}\frac{z^{2}(1+\nu z)\sqrt{1-\nu^{2}}}{[1-3z^{2}-2\nu z^{3}-\gamma^{2}z^{2}(1+\nu z)^{2}]}\;,\label{eq:chargenuz}\eeq
\beq \Phi_{i}=-\sqrt{\frac{5 g_{3}^{2}}{4\pi G_{3}}}\frac{\gamma\nu z^{3}(1+\nu z)}{[1-3z^{2}-2\nu z^{3}-\gamma^{2}z^{2}(1+\nu z)^{2}]}\;.\label{eq:elpotznu}\eeq
At this level, $\Phi_{i}$ is solely a quantum backreaction effect, while it is not apparent the charge too arises from non-vanishing backreaction. 

\subsection*{Temperature}

From the point of view of an observer on the brane, the horizons will emit radiation at their respective (Gibbons-)Hawking temperature, $T_{i}=\frac{\kappa_{i}}{2\pi}$, where $\kappa_i$ is the surface gravity associated to each horizon $r_{i}$. Thus, from surface gravities (\ref{eq:surfgravs}), we find the respective temperatures in terms of $z$ and $\nu$
\beq
    T_i =\frac{|H'(\bar{r}_{i})|}{4\pi}= \frac{z}{2\pi R_3}\left|\frac{2+3\nu z +\nu z^3 (\gamma^{2}(1+\nu z)^{2}-1)}{1-3z^2-2\nu z^3-\gamma^{2} z^2(1+\nu z)^2}\right|\;.
\label{eq:tempallz}\eeq
More specifically, the temperature of the cosmological, outer and inner black hole horizons are, respectively, 
\beq 
\begin{split} 
&T_{c}=\frac{z_{c}}{2\pi R_3}\frac{2+3\nu z_{c} +\nu z_{c}^3 (\gamma^{2}(1+\nu z_{c})^{2}-1)}{3z_{c}^2-1+2\nu z_{c}^3+\gamma^{2} z_{c}^2(1+\nu z_{c})^2}\;,\\
&T_{+}=-\frac{z_{+}}{2\pi R_3}\frac{2+3\nu z_{+} +\nu z_{+}^3 (\gamma^{2}(1+\nu z_{+})^{2}-1)}{3z_{+}^2-1+2\nu z_{+}^3+\gamma^{2} z_{+}^2(1+\nu z_{+})^2}\;,\\
&T_{-}=\frac{z_{-}}{2\pi R_3}\frac{2+3\nu z_{-} +\nu z_{-}^3 (\gamma^{2}(1+\nu z_{-})^{2}-1)}{3z_{-}^2-1+2\nu z_{-}^3+\gamma^{2} z_{-}^2(1+\nu z_{-})^2}\;.
\end{split}
\eeq
Here we introduced $z_{c}=R_{3}/r_{c}x_{1}$, $z_{\pm}=R_{3}/r_{\pm}x_{1}$, and used $r_{-}<r_{+}<r_{c}$, such that $z_{-}>z_{+}>z_{c}$.  Since $r_{+}<r_{c}$, it follows $T_{+}>T_{c}$. In other words, to a static patch observer the dS$_{3}$ black hole is generally not in thermal equilibrium. Situations of thermal equilibrium correspond to situations in which the horizons degenerate. We will return to this shortly.


\subsection*{Entropy and first law}

The four-dimensional bulk black hole has an entropy associated with each of the horizons, given by the Bekenstein-Hawking area-entropy relation
\begin{equation}
\begin{split}
    S^{(4)}_{\text{BH},i} = \frac{\text{Area}(r_i)}{4G_4} &= \frac{2}{4 G_4}\int_0^{2\pi \Delta}d\phi \int_0^{x_1} dx \frac{\ell^2 r_i^2}{(\ell+xr_i)^2} \\
    &= \frac{R_3 \pi}{G_3}\frac{z_i \sqrt{1-\nu^2}}{3z^{2}-1+2\nu z^{3}+\gamma^{2}z^{2}(1+\nu z)^{2}}\;, 
\end{split}
\label{eq:bulkent}\end{equation}
where in the last equality we used $G_{4}=2G_{3}\ell/\sqrt{1-\nu^{2}}$. From the brane perspective, the entropy is a sum of the gravitational entropy (including the higher-derivative contributions) and the matter von Neumann entropy due to the cutoff CFT$_{3}$. That is, the bulk entropy is identified with the generalized entropy, 
\beq S_{\text{BH},i}^{(4)}\equiv S_{\text{gen},i}^{(3)}\;.\label{eq:SBHequalSgen3}\eeq
Since the quantum black hole solution is exact, the generalized entropy (\ref{eq:SBHequalSgen3}) is exact to all orders in $\nu$. To parse out the matter and gravitational contributions to $S_{\text{gen}}^{(3)}$, formally one simply subtracts the gravitational entropy
\beq S_{\text{CFT}}^{(3)}\equiv S_{\text{gen}}^{(3)}-S_{\text{Wald}}^{(3)}\;.\eeq
Here $S_{\text{Wald}}^{(3)}$ is the Wald entropy \cite{Wald:1993nt} for arbitrary diffeomorphism invariant theories of gravity. To leading order in a small-$\ell$ expansion, the gravitational entropy on the brane is simply the three-dimensional Bekenstein-Hawking entropy
\beq S_{\text{BH},i}^{(3)}=\frac{2\pi \bar{r}_{i}}{4G_{3}}=\frac{(1+\nu z)}{\sqrt{1-\nu^{2}}}S_{\text{gen},i}^{(3)}\;.\eeq
Due to its dependence on $\nu$,  $S_{\text{BH}}^{(3)}$ includes quantum backreaction effects. Moreover, since the higher-derivative contributions to the induced brane action (\ref{eq:indactbranecharge}) enter at $\mathcal{O}(\ell^{2})$, they likewise contribute to the generalized entropy starting at order $\mathcal{O}(\nu^{2})$. An expansion of entropy (\ref{eq:bulkent}) in terms of $\nu$ reveals a term linear in $\nu$, which is solely due to the matter entanglement entropy. The analysis of the matter and Wald entropies follow \emph{mutatis mutandis} from the neutral, static quantum BTZ black hole \cite{Emparan:2020znc}, and will therefore not repeat it here.

In summary, the thermodynamic variables of the quantum charged dS$_{3}$ black hole are given by mass (\ref{eq:massz}), charge (\ref{eq:chargenuz}), potential (\ref{eq:elpotznu}), temperature (\ref{eq:tempallz}) and entropy (\ref{eq:bulkent}). It is straightforward to verify that each horizon obeys a first law\footnote{We vary with respect to $z,\nu$ and $\gamma$, keeping other parameters fixed, e.g., $dM=\partial_{z}Mdz+\partial_{\nu}M d\nu+\partial_{\gamma}Md\gamma$.}
\beq dM=T_{i}dS_{\text{gen},i}^{(3)}+\Phi_{i}dQ\;.\label{eq:firstlawcqbh}\eeq
Some comments are in order. First, we see that when accounting for semi-classical backreaction, the classical entropy has been replaced by its generalized counterpart, consistent with the thermodynamics of two-dimensional quantum black holes \cite{Pedraza:2021cvx,Svesko:2022txo}.
Second, the first law holds for all $\nu$. From the brane viewpoint this is a highly non-trivial result, as in principle one would have to compute the generalized entropy, mass, and charge with respect to a resummed version of the induced theory. Due to holography, the bulk automatically performs this resummation. In particular, the classical Bekenstein-Hawking entropy of the bulk black hole exactly computes the entropy of the quantum black hole including the von Neumann entropy of CFT$_{3}$ matter, a highly non-trivial computation from the brane perspective.

Finally, the semi-classical first laws for the black hole and cosmological horizons are
\beq
\begin{split}
    dM &= T_+dS_{\text{gen,}+}^{(3)} + \Phi_+dQ \;, \\
    dM &= -T_-dS_{\text{gen,}-}^{(3)} + \Phi_-dQ\;, \\
    dM &= -T_cdS_{\text{gen,}c}^{(3)} + \Phi_cdQ \;.
    \end{split}
 \label{eq:1stlawC}  \eeq
 The minus sign appearing in the first law for cosmological horizons is standard \cite{Gibbons:1977mu} and can be derived using a quasi-local treatment \cite{Banihashemi:2022htw}. The effect of the minus sign is that adding mass to the static patch leads to a lower cosmological horizon entropy than the entropy of pure de Sitter. This suggests empty de Sitter space is a maximal entropy state with a finite number of degrees of freedom \cite{Banks:2000fe,Banks:2006rx}. We return to this point in Section \ref{sec:entdefs}. 

For illustrative purposes, 
 below we introduce a dimensionless mass parameter $\alpha \in \left[ 0,1 \right]$ and the charge ratio $\rho \in \left[ 0,1 \right]$, defined as
\be \label{alpha_rho}
    M = \alpha M_{\text{N}} + (1-\alpha)M_{\text{E}} \, , \quad \rho \equiv \dfrac{Q}{Q_{\text{u}}} \, .
\ee
Notice small values of $\alpha\in[0,1]$ correspond to black holes with small mass relative to Nariai, while as $\alpha\to1$ the mass $M$ approaches the Nariai system.

 



\subsection*{Thermodynamics of degenerate horizons}

The charged dS$_{3}$ black hole has various limits where the at least two horizons become degenerate. Consequently, the thermodynamics of these limiting geometries is modified relative to their non-degenerate counterparts. 

\vspace{3mm}

\begin{figure}[t!]
  \centering
  \begin{minipage}[t]{0.47\textwidth}
    \includegraphics[width=.9\textwidth]{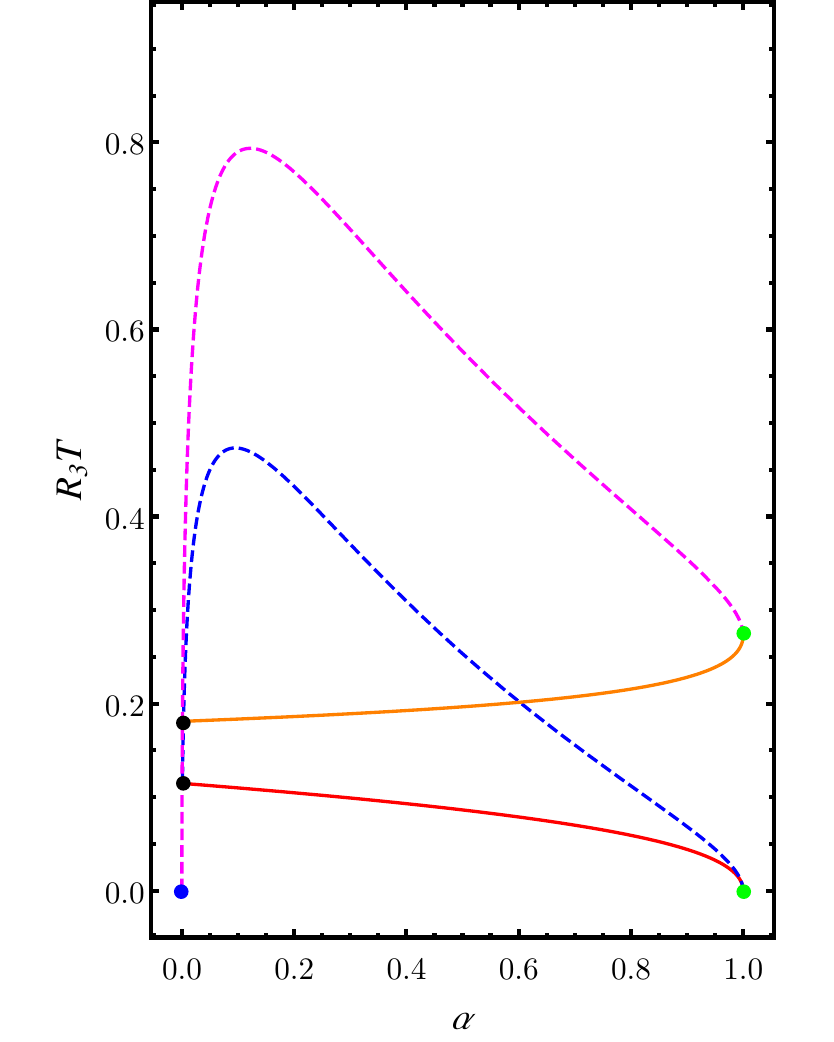}
  \end{minipage}
  \hfill
  \begin{minipage}[t]{0.47\textwidth}
    \includegraphics[width=.9\textwidth]{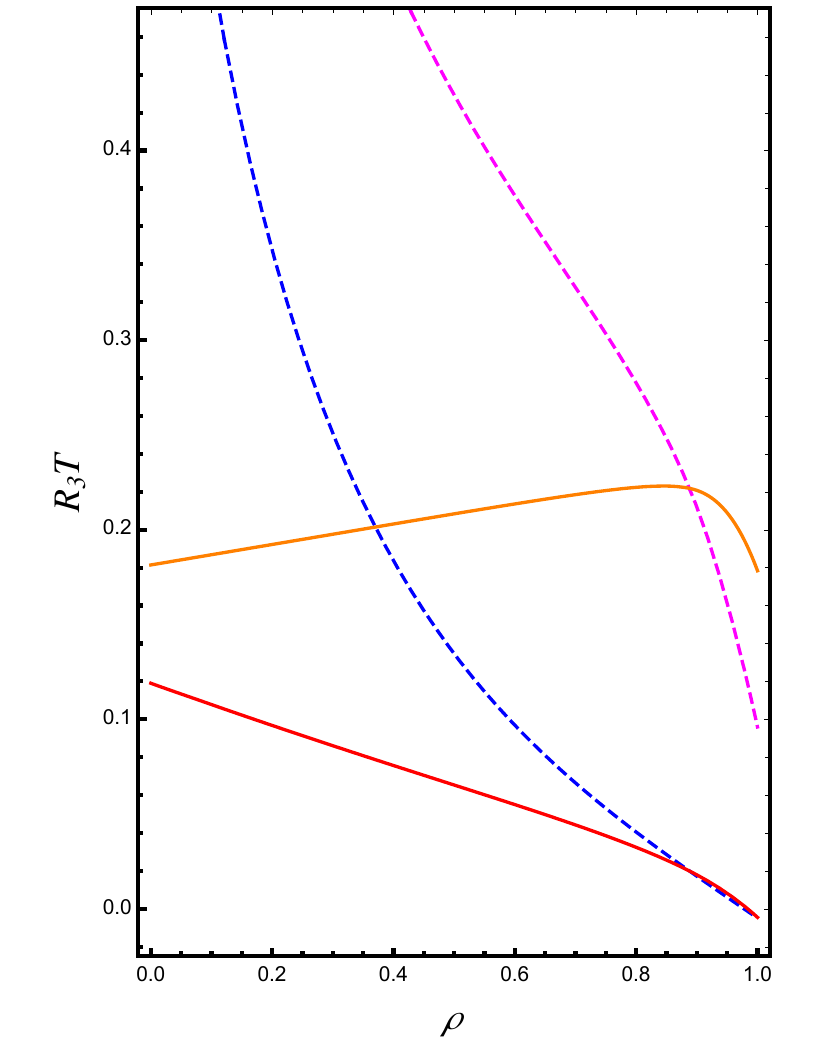}
  \end{minipage}
  \caption{\small  \textbf{Left:} Horizon temperature $T$ as a function of the dimensionless parameter $\alpha$ \eqref{alpha_rho} at $\rho = 0.35$ and $\nu = 1/3$. Dashed curves correspond to the outer black hole horizon temperatures (upper, magenta curve refers to temperature with Bousso-Hawking normalization), while solid curves denote cosmological horizon temperatures (upper, orange curve is in Bouss-Hawking normalization).
  The black (left), green (right) and blue (lower, left) dots correspond to the lukewarm, Nariai and extremal limits, respectively. \textbf{Right:} Horizon temperature as a function of the dimensionless charge parameter $\rho$ at $\alpha = 0.50$ and $\nu = 1/3$. This plot has the same coloring scheme as the left plot.}
  \label{fig:T}
\end{figure}

\noindent \emph{Extremal black hole:} Geometrically, the extremal black hole occurs when the inner and outer black hole horizons coincide, $\bar{r}_{+}=\bar{r}_{-}$. Thermodynamically, the extremal black hole has vanishing temperature $T_{+}=T_{-}=0$, made apparent from the surface gravities (\ref{eq:surfgravs}), while the cosmological horizon temperature is non-zero, 
\beq T_{c}=-\frac{\kappa_{c}}{2\pi}=-\frac{(\bar{r}_{c}-\bar{r}_{+})^{2}(\bar{r}_{c}+\bar{r}_{+})}{2\pi\bar{r}_{c}^{2}R_{3}^{2}}\;.\label{eq:cosmoTatext}\eeq
The entropy of extremal black holes $S_{\text{E}}$, derived from the extremal four-dimensional area \eqref{eq:SBHequalSgen3} at $z\equiv z_{\text{extremal}}$, remains non-zero. See Figure \ref{fig:T}. This effect is due to the dominance of quantum gravitational fluctuations of the AdS$_2$ throat \eqref{eq:extAdS2} at sufficiently low temperatures. The quantum effects of the CFT dominate away from the throat. However, we can still study how the semi-classical description of the throat is expected to break down when backreaction is included. In general, $S_{\text{E}}$ is a very complicated expression depending on $q$ and $\nu$. In order to gain some insight of its behavior we will study the limit in which $q$ is small, resulting in
\be
    S_{\text{E}} = \dfrac{R_3 \pi}{2G_3} \nu q^2 \sqrt{1+\nu^2} + \mathcal{O}(q^3) \, .
\ee
Note the extremal entropy vanishes as either $q$ or $\nu$ approach zero, as in either case the extremal limit does not exist.

\vspace{3mm}

\noindent \emph{Charged Nariai black hole:} The Nariai black hole occurs when the outer black hole and cosmological horizons coincide, $\bar{r}_{c}=\bar{r}_{+}=\bar{r}_{\text{N}}$. Naively, the temperature $T_{\text{N}}$ of the charged Nariai solution is vanishing. In the case of Schwarzschild-de Sitter, Bousso and Hawking argued the Nariai temperature is in fact non-zero, and the vanishing of the temperature is a consequence of using a less natural choice of normalization of the time-translation Killing vector \cite{Bousso:1996au}.\footnote{Further, the near-horizon Nariai geometry is  $\text{dS}_{2}\times S^{2}$, with the temperature of $\text{dS}_{2}$ static patch observer.} Technically, the Bousso-Hawking normalization is chosen such that the time-translation Killing vector $\bar{\zeta}^2 =-1$ at the radius $\bar r_0$ where the blackening factor $H(\bar r)$ obtains a maximum, i.e., when $H'(\bar{r}_{0})=0$. Physically, the radius $\bar{r}_{0}$ corresponds to the location where an observer can stay in place without accelerating, consistent with the asymptotically flat Schwarzschild black hole or empty de Sitter where $\bar{r}_{0}\to0$. 
Consequently, the horizon generating Killing vector is $\bar{\zeta}_{\bar{t}} = \partial_{\bar{t}}/\sqrt{H(\bar{r}_0)}$ and the temperatures $\bar{T}_i$ defined with respect to the Bousso-Hawking choice of normalization  are\footnote{Note that this choice of normalization does not change that the extremal black hole temperature is zero. Alternatively, had a similar choice of normalization been chosen for the extremal black hole, i.e., normalize the time-translation Killing vector with respect to the positive root of $H'(\bar{r})=0$ between $r_{-}$ and $r_{+}$, then the extremal black hole would have non-vanishing temperature and Nariai would have zero temperature.} 
\beq
    \bar{T}_i = \dfrac{T_i}{\sqrt{H(\bar{r}_0)}} \, .
\eeq
Exact expressions for the Bousso-Hawking temperatures $\bar{T}_{+,c}$ are cumbersome, however, from Figure \ref{fig:T} we see the temperatures equal the same non-zero finite value in the Nariai limit. Another physical consequence of the Bousso-Hawking normalization, however, is that the electric potential (\ref{eq:chemicalpotentialbulk}) is rescaled such that $\Phi_{i}$ diverges in the Nariai limit, as in \cite{Castro:2022cuo}.

\vspace{3mm}

\noindent \emph{Ultracold black hole:} The ultracold black hole occurs when all horizon radii coincide, $\bar{r}_{-}=\bar{r}_{+}=\bar{r}_{c}=\bar{r}_{\text{u}}$. In this case, the temperature vanishes, $T_{\text{u}}=0$, evident from (\ref{eq:cosmoTatext}). 

\vspace{3mm}

\noindent \emph{Lukewarm black hole:} Another configuration in which the two temperature match away from the Nariai limit, is the ‘lukewarm' solution \cite{Romans:1991nq}. The geometry is non-singular and the two horizons are in thermal equilibrium, $T_{+}=T_{c}$ (see Figure \ref{fig:T}).

\section{Schottky peaks of quantum dS black holes} \label{sec:schotpeaks}

A thermal system's heat capacity 
 encodes useful information about its phase behavior and underlying microscopic description. For example, first and second order phase transitions are often accompanied by divergences or discontinuities in the heat capacity. 
 Moreover, a ``peak'' in the  heat capacity, known as the Schottky anomaly, naturally occurs in a system with a maximum number of available energy levels. A prototypical example is that of a two-level system with a zero energy ground state with energy $E_{0}=0$ and an excited state $E_{1}=\epsilon$. The average energy of the system is 
 \beq U=\frac{\epsilon e^{-\beta \epsilon}}{1+e^{-\beta\epsilon}}\;,\eeq
 for inverse temperature $\beta=T^{-1}$. At large temperature ($T\gg\epsilon$), the average energy saturates at $U(\beta\to0)=\epsilon/2$, and vanishes at low temperatures $T\ll\epsilon$ (as does the entropy $S=\log(1+e^{-\beta\epsilon})$). The heat capacity, meanwhile,
 \beq C\equiv\left(\frac{\partial U}{\partial T}\right)_{\hspace{-1mm}\epsilon}=(\beta\epsilon)^{2}\frac{e^{\beta\epsilon}}{(1+e^{\beta\epsilon})^{2}}=(\beta \epsilon)^{2}\sum_{n=1}^{\infty}(-1)^{n-1}ne^{-n\beta\epsilon}\;\label{eq:heatcapSchot}\eeq
 has a peak at $T\approx .417\epsilon$ and vanishes at high and low temperatures.\footnote{Schottky peaks also arise in truncated systems. Consider, for example the quantum harmonic oscillator with energy levels $E_{n}=\epsilon(n+\frac{1}{2})$. The heat capacity is
 $$C=(\beta \epsilon)^{2}\frac{e^{\beta\epsilon}}{(1-e^{\beta\epsilon})^{2}}=(\beta \epsilon)^{2}\sum_{n=1}^{\infty}ne^{-n\beta\epsilon}\;.$$
 Truncating to finite $n=N$, thereby restricting the space of states to be occupied, results in a peak similar to the one observed for the two-level system (\ref{eq:heatcapSchot}) \cite{Johnson:2019ayc}.}
Qualitatively, the peak arises because at low temperature, small-$T$ fluctuations hardly change the internal energy of the system to push the system to its excited state, while for high-temperatures the energy of the system has reached its maximum value such that further increasing $T$ does not bring the system to occupy other states. This makes heat capacity a powerful tool for probing the microscopic nature of a thermal system's macrostates.

Black holes in de Sitter space are also examples of thermal systems with a state of maximum energy (the Nariai limit). It is natural then to explore their heat capacity in search for Schottky-like behavior. Indeed, it was found that the heat capacity of classical dS black holes feature (inverted) Schottky peaks \cite{Dinsmore:2019elr,Johnson:2019ayc}, and thus, microscopically, a dS black hole can be viewed as a system with a finite number of energy levels.\footnote{AdS black holes feature Schottky behavior when in an ensemble of fixed ``thermodynamic volume'' \cite{Johnson:2019vqf}.} Likewise, below we will show that quantum dS$_{3}$ black holes (focusing on the neutral and charged static geometries) have a Schottky behavior. In fact, while here we focus we on dS quantum black holes, as we discuss in Section \ref{sec:disc}, all known exact quantum black holes have mass confined to a finite range, and thus all quantum black holes should have Schottky peaks.


Before discussing the charged quantum dS$_{3}$ black hole, it behooves us to first review the classical RN-de-Sitter black hole \cite{Dinsmore:2019elr,Johnson:2019ayc}. We give a complete analytic treatment, uncovering aspects not reported in the literature. Succinctly, the key observations are:
\begin{itemize}
    \item An inverted Schottky peak is always present for the neutral Schwarzschild-dS$_4$ solution, when $T_S \sim 1/R_4$. Much like for systems with a maximum energy level, the heat capacity vanishes at $T=0$ and $T\to \infty$ (where here $T$ does not assume the Bousso-Hawking normalization). 
    \item For RN-dS$_{4}$ with charge $Q$, a Schottky anomaly is only present if $Q/Q_u \lessapprox 0.38$, and always in the Nariai branch. Further, charged black hole features a maximum temperature $T_{\text{max}}$, where the heat capacity diverges. In this case, the behavior of the heat capacity is better represented by a system whose energy spectrum features a large gap, after which many energy levels are again available.
    \item Quantum dS black holes share most of the qualitative traits as the classical RN-dS$_{4}$, for an appropriate range of backreaction $\nu$. 
\end{itemize}


\subsection{Classical Reissner-Nordstr{\"o}m-dS$_4$}

The metric for an electrically charged de Sitter$_{4}$ black hole  is given by
\beq
\label{eq:RN_dS}
ds^2 = -\left(1-\frac{r^2}{R_4^2} -\frac{2G_4 M}{r}+\frac{Q^2}{r^2}\right)dt^2 +\left(1-\frac{r^2}{R_4^2} -\frac{2G_4 M}{r}+\frac{Q^2}{r^2}\right)^{-1}dr^2+ r^2 d\Omega_2 \ .
\eeq
The entropy $S_{+}$ and temperature $T_{+}$ associated with the outer black hole horizon $r=r_{+}$ are
\beq
\label{eq:RN_temp}
T_+ = \frac{1}{4\pi r_+} \left(1-3 \frac{r_+^2}{R_4^2}-\frac{Q^2}{r_+^2} \right)
\eeq
and
\beq
S_+ = \frac{4\pi r_+^2}{4G_{4}} \ .
\eeq
Note that the range of $r_+$ spans from the extremal radius $r_{e}$ to the Nariai radius $r_{\text{N}}$, and we do not assume a Bousso-Hawking normalization such that temperature (\ref{eq:RN_temp}) vanishes at both the extremal and Nariai limits.
The system reaches a maximum temperature at
\beq
r_{T_\text{max}} = \frac{R_4}{\sqrt{6}}\sqrt{-1+\sqrt{1+36\frac{Q^2}{R_4^2}}} \ ,
\eeq
which disappears in the neutral limit.

\begin{figure}[t!]
  \centering
  \begin{minipage}[t]{0.47\textwidth}
    \includegraphics[width=\textwidth]{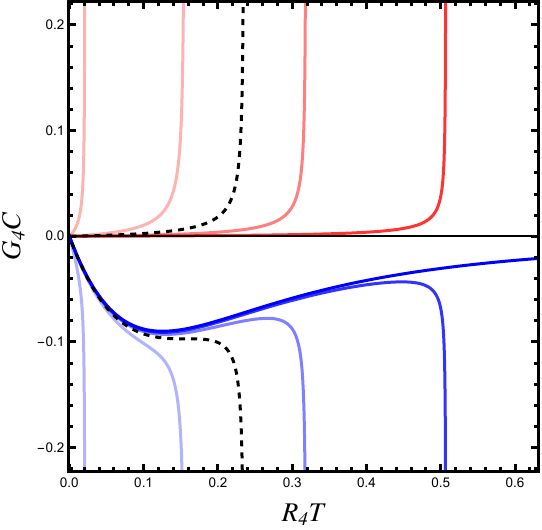}
  \end{minipage}
  \hfill
  \begin{minipage}[t]{0.47\textwidth}
    \includegraphics[width=\textwidth]{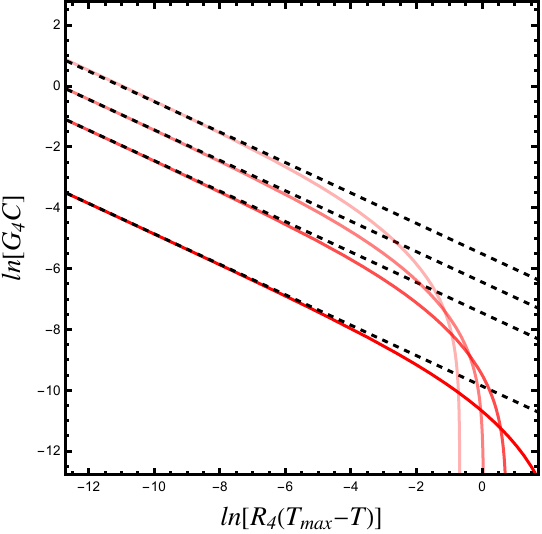}
  \end{minipage}
  \caption{\small  \textbf{Left:} Constant charge heat capacity $C$ against temperature $T$ of the (3+1)-dimensional R-N dS black hole with unit dS radius $R_4=1$. We depict the extremal branch in red and the Nariai branch in blue. From darkest to lightest, the solid curves correspond to black holes with charge $Q/Q_u = 0,0.2,0.3,0.5,0.9$, where $Q_u$ is the ultracold mass. The dashed line corresponds to $Q/Q_u=0.38$, the maximum charge for which the Schottky peak is present. ìNote that the $Q=0$ branch does not feature the extremal branch. The asymptote corresponds to the maximum temperature of the black hole $T_\text{max}$.  \textbf{Right:} Critical behavior of $C$ of the RN black hole close to the maximum temperature $T_\text{max}$ as a function of temperature. From darkest to lightest, the curves correspond to black holes with charge $Q/Q_u = 0.01,0.05,0.1,0.2$.}
  \label{fig:RN_schottky}
\end{figure}

From the first law, the heat capacity at constant charge is
\begin{equation}
    C = \left. \frac{\partial M}{\partial T_+} \right|_{Q} = \left. T_+ \frac{\partial S_+}{\partial T_+} \right|_{Q} \ .
\end{equation}
Given both $S_+$ and $T_+$ at constant charge are function solely of $r_+$, we can express $C$ as a parametric function of the outer horizon radius
\beq
\label{eq:CP_RN}
C = \frac{\pi r_+^2}{2G_4} \frac{3 r_+^4- R_4^2 r_+^2+ Q^2 R_4^2}{3 r_+^4+R_4^2 r_+^2-3Q^2 R_4^2} \ .
\eeq
See Figure \ref{fig:RN_schottky} for an illustration.
Since the temperature (\ref{eq:RN_temp}) in the physical range of $r_+$ is an everywhere non-negative and continuous function, we can identify two different branches in the heat capacity (\ref{eq:CP_RN}) for each value of $T_+$. Namely: (i) \textit{extremal branch}  black holes for which $r_{e}<r_{+}<r_{T_{\text{max}}}$, and (ii) \emph{Nariai branch} black holes with $r_{T_\text{max}}<r_+\leq r_N$. 
Note that the extremal branch is present only for non-zero charge and has positive heat capacity (implying thermal stability), contrary to the Nariai branch. 

Next, from the root structure of the denominator in heat capacity \eqref{eq:CP_RN}, we see $C$ diverges when $r_+ = r_{T_\text{max}}$, i.e., at the maximum temperature of the black hole. The divergence is a Davies curve \cite{davies1977}, corresponding to the phase transition from a thermodynamically stable state to one that is unstable. In fact, we can easily determine the critical exponent of the divergence. Since for $Q \neq 0$ there are two distinct real roots (and two imaginary roots) for $r_+$, the denominator can be trivially rewritten near the pole as
\beq
 3 r_+^4+R_4^2 r_+^2-3Q^2 R_4^2 \approx (r_+-r_{T_\text{max}})  [12 r_{T_\text{max}}^3 + 2R_4^2 r_{T_\text{max}}] \ .
\label{eq:denoexp}\eeq
Similarly, we can expand temperature \eqref{eq:RN_temp} near its maximum  and invert for the radius. Picking the extremal branch for simplicity (although the critical exponent turns out to be equivalent for both branches), we find
\beq
r_+ - r_{T_\text{max}} \approx  - \left[4\pi r_{T_\text{max}}^5 \frac{T_\text{max}-T}{6 Q^2-1 r_{T_\text{max}}^2}\right]^{1/2} \ .
\label{eq:rpminusrt}\eeq
Substituting (\ref{eq:denoexp}) and (\ref{eq:rpminusrt}) into  the heat capacity \eqref{eq:CP_RN} we find that close to the maximum temperature,
\beq
C \approx -\frac{\pi r_{T_\text{max}}^2}{2G_4} \left(\frac{3 r_{T_\text{max}}^4 - R_4^2 r_{T_\text{max}}^2 +Q^2 R_4^2}{12 r_{T_\text{max}}^3 + 2 R_4^2 r_{T_\text{max}}} \right) \sqrt{\frac{6 Q^2- r_{T_\text{max}}^2}{4\pi r_{T_\text{max}}^5}}  \ (T_{\max}- T)^{-1/2} \,, 
\eeq
such that the critical exponent is $\alpha=1/2$, following the convention  $C \propto |T_c- T|^{-\alpha}$. 


For a wide range of charge the Nariai branch features a Schottky anomaly, as illustrated in Figure \ref{fig:RN_schottky}. The anomaly is always present in the neutral black hole, but increasing the charge pushes the maximum temperature towards $T=0$, eventually hiding the local minimum of $C$. In order to establish the parameter range for which a Schottky anomaly is present, it is necessary to look at the first derivative of the heat capacity:
\beq
\label{eq:CP_derRN}
\frac{\partial C}{\partial T_+} = 8\pi R_4^2 r_+^5 \frac{3Q^4 R_4^4 -6 Q^2R_4^4 r_+^2+ R_4^2 (30Q^2+R_4^2)r_+^4 -6 R_4^2 r_+^6-9r_+^8}{(-3Q^2 R_4^2+R_4^2 r_+^2+3r_+^4)^3} \ .
\eeq
The sole root of the denominator in the physical range of $r_+$ is $r_{T_\text{max}}$. Therefore, it suffices to find zeros of the numerator, though their explicit expressions are not particularly illuminating.
Nonetheless, we can easily check that the neutral black hole always features a Schottky anomaly at
\beq
r_{\text{S}}= R_4 \sqrt{\frac{\sqrt{2}-1}{3}} \approx 0.64 r_{\text{N},Q=0} \ ,
\eeq
where $r_{\text{N},Q=0}$ is the neutral Nariai radius. The corresponding Schottky temperature is
\beq
T_{\text{S}} = \frac{1}{4\pi R_4}\frac{2\sqrt{3}-\sqrt{6}}{\sqrt{\sqrt{2}-1}} \approx \frac{1.68}{4\pi R_4}\ .
\eeq

Adding charge significantly complicates the picture. Other than the unphysical $r_+=0$ root, the existence of a local extremum relies on the existence of a positive root of the numerator in the derivative \eqref{eq:CP_derRN}. Rewriting it in terms of the normalized charge $\rho=Q/Q_u$, where $Q_u= R_4/(2\sqrt{3})$ is the ultracold charge, the numerator reads
\beq
N(r_s) = - 9 r_s^4
- 6 R_4^2 r_s^3+ R_4^4 r_s^2\left( 1+\frac{5}{2}\rho^2\right)-\frac{1}{2}R_4^6 \rho^2 r_s +\frac{1}{48}R_4^8 \rho^4 \ ,
\eeq
where $r_s \equiv r_+^2$. Not only the existence of positive $r_s$ roots of $N(r_s)$ is required, but also $r_E \leq r_s \leq r_N$. Observing that the heat capacity vanishes at $r_+ = 0$ and $r_+=r_E$, we already know (for $Q\neq 0$) there is always an extremum in that unphysical range. Further, $N(0)>0$ and $N(\pm \infty) = -\infty$, meaning there always exists at least one root in the unphysical $r_s <0$ half-line. Descartes' rule of signs shows such a root is unique. Therefore, if two other roots exist, they correspond to the pair of local minimum/maximum that is required for the Schottky anomaly in the Nariai branch. The condition for its presence is thus reduced to a positivity condition on the discriminant of the quartic equation $N(r_s)=0$:
\beq
\Delta_\text{N} = -3 R_4^{24} \rho^4 ( -4 + 28 \rho^2 + 9\rho^4-82 \rho^6+49 \rho^8) > 0\  .
\eeq
Introducing $\rho_s \equiv \rho^2 $, it can be easily checked that $\rho_s=1$ is a root of $\Delta_\text{N}=0$. The latter quartic equation has a positive determinant, meaning that it has four real roots in terms of $\rho_s$; the two positive of which produce two pairs of symmetric roots about $\rho=0$. Since at small positive $\rho$ the discriminant $\Delta_\text{N}>0$, it is the smallest positive root $\rho^*$ that provides the maximum normalized charge that a RN-de Sitter black hole can have if it is to feature a Schottky anomaly. It is straightforward to solve the quartic equation in terms of $\rho_s$ to find
\beq
\rho^* \approx 0.38 \ .
\eeq
The limiting solution is shown as a dashed line in the right panel of Figure \ref{fig:RN_schottky}.

While here we focused on charged, non-rotating black holes, a similar analysis can be carried out for the Kerr-de Sitter black hole, as investigated numerically in \cite{Johnson:2019ayc}. 


\begin{figure}[t!]
  \centering
  \begin{minipage}[t]{0.47\textwidth}
    \includegraphics[width=\textwidth]{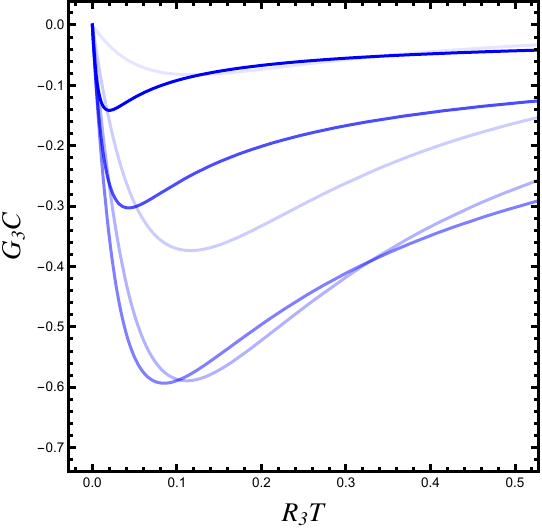}
  \end{minipage}
  \hfill
  \begin{minipage}[t]{0.47\textwidth}
    \includegraphics[width=\textwidth]{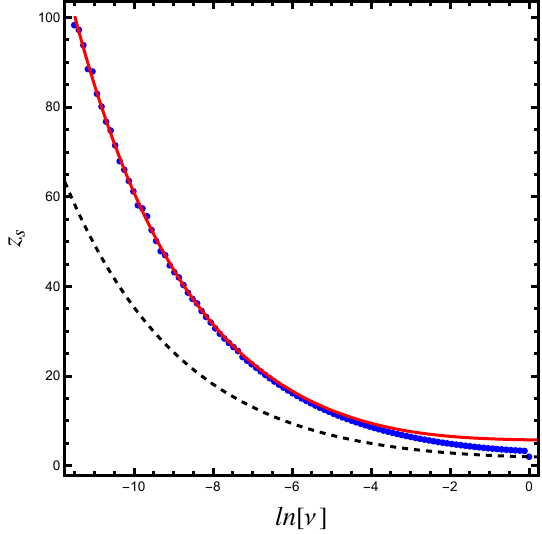}
  \end{minipage}
  \caption{\small  \textbf{Left:} Heat capacity $C$ for the neutral qSdS$_3$ black hole as a function of temperature $T$. From darkest to lightest, the curves correspond to black holes with backreaction parameter $\nu = 0.001,0.01,0.1,0.5,0.8,0.99$. \textbf{Right:} Value of the $z$ parameter for which $C$ is a minimum as a function of $\nu$, identified numerically. The black dashed line corresponds to the lower bound for the $z$ parameter $z_N$, whilst the red curve is the analytic approximation for small backreaction.}
  \label{fig:neutral_qSdS}
\end{figure}

\subsection{Schottky anomalies in quantum dS black holes}
Schottky anomalies appear also for quantum corrected 2+1-dimensional de Sitter black holes. 
An analytic treatment is more cumbersome, however, so we will mainly resort to numerical analysis to study how backreaction modifies the Schottky behavior. 

We begin with the simpler neutral quantum black hole. 
The heat capacity is\footnote{Upon Wick rotation $R_{3}\to -i\ell_{3}$ such that $z^{2}\to-z^{2}, \nu^{2}\to-\nu^{2}$ and $\nu z\to \nu z$, we recover the heat capacity of the neutral quantum BTZ black hole \cite{Frassino:2023wpc}.}
\begin{equation}
\label{eq:cp_qSdS}
    C =\frac{\partial M}{\partial T}= -\frac{R_3}{8G_3}\frac{4\pi z \sqrt{1-\nu^2} (-2-3z\nu+z^3\nu)(1+3 z^2+12 z^3 \nu)}{(1+z^3 \nu)(-1+3z^2+2 z^3 \nu)(1+3z^2+3 z \nu +z^3\nu)} \ ,
\end{equation}
and is illustrated in Figure \ref{fig:neutral_qSdS}. 
As before, a Schottky anomaly is present for the entire allowed physical range of the backreaction parameter $0<\nu<1$.
Differentiating \eqref{eq:cp_qSdS} with respect to temperature yields a rational function whose numerator is high-ordered a polynomial such that identifying the location of the Schottky peak in terms of $z_{\text{S}}$ cannot be determined analytically for arbitrary $\nu$.
Progress can be made in the regime of small backreaction, where we keep only terms quadratic in $\nu$, 
\begin{equation}
\begin{split}
    -2 (1 + 3 z^2)^3 &- 
 2 (3 + 68 z^2 + 54 z^4 + 108 z^6 + 135 z^8) z \nu \\
 &+ 
  (1 - 222 z^4 + 25 z^6 - 594 z^8 - 837 z^{10}+ 
    27 z^{12}) \nu^2 = 0 \;.
\end{split}
\end{equation}
Solving to find $z_{\text{S}}$ is prohibitive, however, we can flip the problem on its head and instead determine the value of $\nu$ such that the minimum heat capacity is attained for a black hole with $z=z_{\text{S}}$ by readily solving the quadratic equation for $\nu$ and imposing $z \geq z_\text{N}$ and $0\leq \nu \leq 1$. The analytic expression is cumbersome, but we show in Figure \ref{fig:neutral_qSdS} how the small-backreaction approximation performs against the numerical solution.

\begin{figure}[t!]
\centerline{\includegraphics[width=0.5\textwidth]{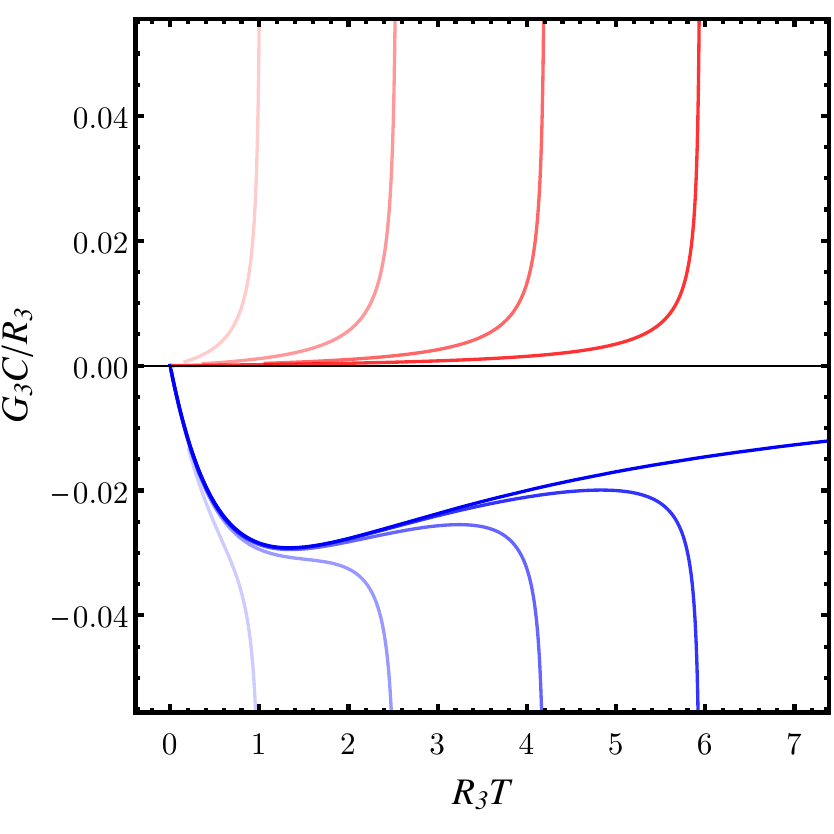}}		\caption{Heat capacity $C$ of the charged qSdS black hole as a function of temperature for $\nu=1/3$. The curves with lighter colors correspond to higher values of $\rho$. Two branches are recognizable. The red line starts from the extremal black hole at $T=0$ and covers all the solutions up to a turnaround temperature $T_{\text{max}}$, whilst the blue line covers the remaining family of solutions up to the Nariai black hole. 
}
\label{fig:Schottky}\end{figure}

When charge is introduced the picture is enriched. Much like in the higher-dimensional case, a second branch with positive heat capacity appears, corresponding to black holes closer to the extremal solution. Moreover, the temperature is now bounded from above by a maximum temperature $T_{\text{max}}$.
As observed in Figure \ref{fig:Schottky}, the heat capacity diverges near as $T$ approaches the maximum temperature. 

\begin{figure}[t!]
  \centering
  \begin{minipage}[t]{0.47\textwidth}
    \includegraphics[width=\textwidth]{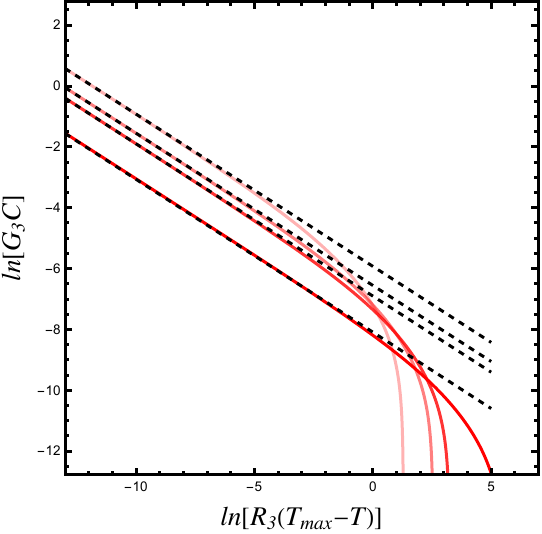}
  \end{minipage}
  \hfill
  \begin{minipage}[t]{0.47\textwidth}
    \includegraphics[width=\textwidth]{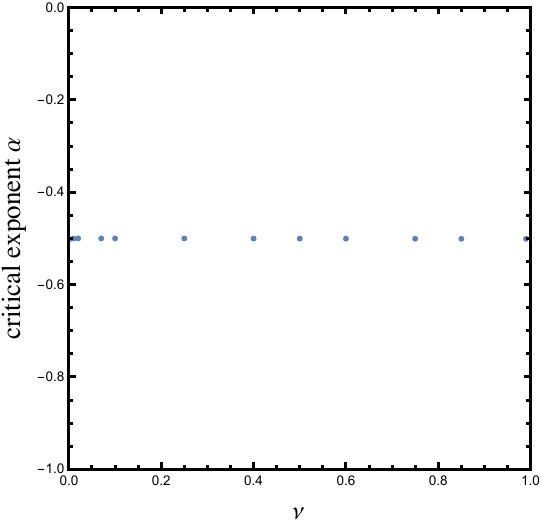}
  \end{minipage}
  \caption{\small  \textbf{Left:} Critical behavior of $C$ of the quantum corrected charged black hole with $q=0.1$ close to the maximum temperature $T_\text{max}$ as a function of $T$. From darkest to lightest, the curves correspond to backreaction parameter $\nu = 0.001,0.01,0.02,0.07$. The black dashed curves are the linear fit for the critical behavior in log-log space. \textbf{Right:} Best-fit value for the critical exponent for the asymptotic behavior of $C$ against $T$ close to the maximum temperature.}
  \label{fig:qRN_schottky}
\end{figure}

Performing the same analysis as in the previous section to extract the critical exponent $\alpha$, we expect $\alpha$ to be a postitive rational number, independent of $\nu$. 
To see this, note that, close to the maximum temperature, $T$ as a function of $z$ will always behave as $z^n$, where $n\geq 2$. In practice, if $z_{\text{max}}$ is not a root of the second derivative of the temperature,
then $\partial T/\partial z$ is a rational function whose numerator is a ninth degree polynomial. 
Close to the maximum temperature, $C \sim (z-z_{\text{max}})^{-p}$, where $p$ is the degeneracy of the root. Therefore, the asymptotic behavior of the heat capacity is of the form
\begin{equation}
    C \propto (T_{\max}-T)^{-p/n} \ ,
\end{equation}
where both $p$ and $n$ are integers. The charge $q$ and backreaction parameter $\nu$ can affect the multiplicity of the roots, discontinuously changing the critical exponent. In Figure \ref{fig:qRN_schottky} we numerically extract (at fixed $q$) the critical exponent for a range of backreaction parameters $\nu$, finding that the quantum black holes have $\alpha=1/2$, as in RN-dS$_{4}$. Further, the Schottky anomaly is present only in the Nariai branch, and above a critical charge $\rho^*$, the peak disappears.

\section{Nucleation of charged quantum-dS$_{3}$ black holes}\label{sec:entdefs}


In the absence of backreaction, empty de Sitter space corresponds to a state of (finite) maximum entropy \cite{Banks:2000fe,Banks:2006rx}. Meanwhile, excited states, e.g., massive particles or black holes in asymptotic dS, have lower entropy, the smallest such entropy state being the Nariai solution. Such configurations can thus be thought of as constraining the degrees of freedom of empty de Sitter space. Likewise, accounting for semi-classical backreaction, (quantum) dS$_{3}$ is a state of maximal entropy whilst the quantum Schwarzschild-dS$_{3}$ black hole is a constrained state \cite{Emparan:2022ijy}. In either context, the probability to create such constrained states in the de Sitter static patch is controlled by entropy deficits, the entropy difference between empty dS and its excited states. More precisely, the probability $\Gamma$ for a fluctuation to an excited state with entropy $S_{ex}\leq S_{\text{dS}}$ is
\beq \Gamma \sim e^{-\Delta S}\;,\label{eq:nucprobv1}\eeq
with entropy deficit $\Delta S\equiv S_{\text{dS}}-S_{ex}$. The entropy deficit controlling the nucleation of a quantum SdS$_{3}$ black hole takes the same form, where classical entropies are replaced by their generalized counterpart, i.e., $\Delta S=S_{\text{gen},\text{dS}}-S_{\text{gen},ex}$ \cite{Emparan:2022ijy}. 

Euclidean path integral methods can be used to derive the rate (\ref{eq:nucprobv1}) of nucleating black holes in de Sitter of arbitrary mass and charge \cite{Mann:1995vb,Hawking:1995ap}. Following \cite{Morvan:2022ybp,Morvan:2022aon}, this is formally accomplished via a saddle-point approximation of the following Euclidean path integral 
\beq \int \hspace{-1mm} D\Sigma\,\Psi_{\Sigma}^{\dagger}\Psi_{\Sigma}=\int \hspace{-1mm} Dg DA \, e^{-I_{E}[g,A]}\;.\label{eq:nucratearbv2}\eeq
Here, $\Psi_{\Sigma}$ is the no boundary Hartle-Hawking wavefunction \cite{Hartle:1983ai} where $\Sigma$ is the boundary of a Euclidean manifold labeling different final states; the pair creation rate $\Gamma_{\Sigma}$ of a black hole of \emph{fixed} mass and charge is approximately given by the square $\Gamma_{\Sigma}=\Psi^{\dagger}_{\Sigma}\Psi_{\Sigma}$. Thus, the nucleation rate of a black hole of arbitrary mass and charge is given by the integral of the modulus of the Hartle-Hawking wavefunction over the complete set of final states. Finally, $I_{E}[g,A]$ denotes the (off-shell) Euclidean action on the entire compact Euclidean manifold endowed with Euclidean metric $g$ and gauge field $A$.

To proceed, one implements a semi-classical saddle point analysis, comparing the on-shell actions for the (Euclidean) charged dS black hole and empty de Sitter space. The Euclidean geometry is found by Wick rotating the time coordinate $t\to it=t_{E}$ of the Lorentzian metric (\ref{eq:qsdsc}).  The Euclidean time coordinate $t_{E}$ is periodically identified with a real parameter $\beta$ such that $\beta$ is fixed to be the inverse temperature characterizing a thermal canonical ensemble. For arbitrary $\beta$ the Euclidean geometry is regular everywhere except at the black hole and cosmological horizon, where there exist conical singularities.\footnote{Generally, only one of the conical singularities can be removed via an appropriate choice of $\beta$, but not both (special exceptions include the extremal, Nariai, ultracold and lukewarm, solutions which have a single conical singularity). This manifests in dS black holes being out of thermal equilibrium systems.} Thus, away from the conical singularities, the Euclidean charged quantum dS black hole is a regular solution to the Euclidean gravitational field equations on the brane. At the Euclidean horizons, however, the conical singularities give rise to divergences in curvature quantities, e.g., a delta-function singularity in the Ricci scalar at each horizon (cf. \cite{Fursaev:1995ef}). 

From the brane perspective, evaluating the on-shell Euclidean action is cumbersome to accomplish in practice due to the infinite tower of higher-derivative terms and not knowing the explicit form of $I_{\text{CFT}}$. Instead, it behooves us to instead evaluate the classical bulk on-shell action of Einstein-Maxwell coupled to a brane action. In the absence of a brane, following  analogous logic presented in \cite{Morvan:2022ybp,Morvan:2022aon}, the action of the Euclidean solution is equal to the sum of classical Bekenstein-Hawking entropies of the black hole and cosmological horizon \cite{Emparan:2000fn}
\beq I_{E}^{\text{on-shell}}=-(S_{\text{BH},+}^{(4)}+S_{\text{BH},c}^{(4)})\;.\label{eq:Ionshell4d}\eeq
Including the brane, it is expected that the on-shell action for quantum dS$_{3}$ braneworld black holes be equal to the sum of the generalized entropies with respect to each horizon \cite{Emparan:2022ijy}
\beq I_{E}^{\text{on-shell}}=-(S_{\text{gen},+}^{(3)}+S^{(3)}_{\text{gen},c})\;.\label{eq:Ionshell3d}\eeq
While the precise derivation of this statement is beyond the scope of this article, the logic is straightforward. Recall the on-shell action of the four-dimensional geometry dual to the quantum BTZ black hole \cite{Kudoh:2004ub} gives rise to the same horizon thermodynamics. Performing the analysis, \emph{mutatis mutandis}, for the bulk geometry dual to the charged quantum dS$_{3}$ black hole at arbitrary $\beta$ would result in (\ref{eq:Ionshell4d}), and (\ref{eq:Ionshell3d}) directly follows from the identification of bulk and brane entropies (\ref{eq:SBHequalSgen3}).

\begin{figure}[t!]
  \centering
  \begin{minipage}[t]{0.47\textwidth}
    \includegraphics[width=.9\textwidth]{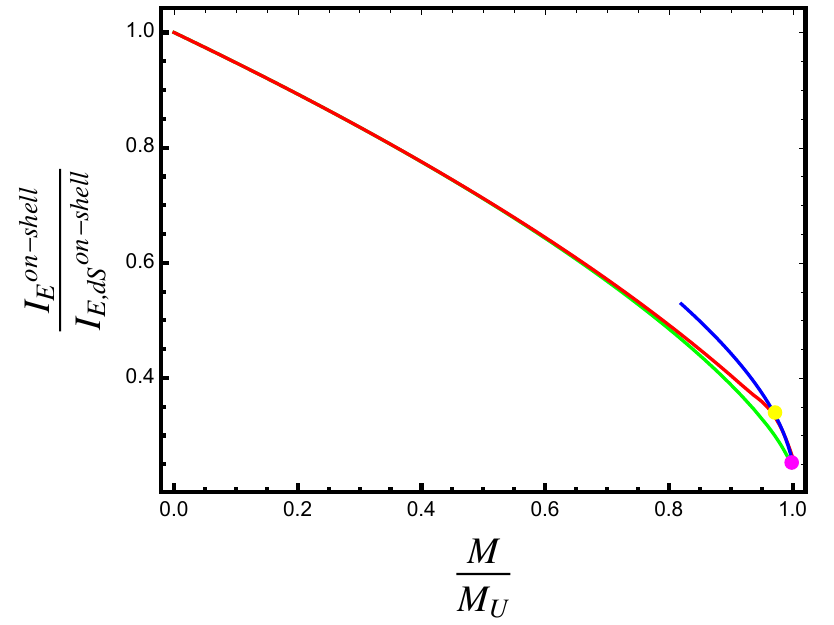}
  \end{minipage}
  \hfill
  \begin{minipage}[t]{0.47\textwidth}
    \includegraphics[width=.9\textwidth]{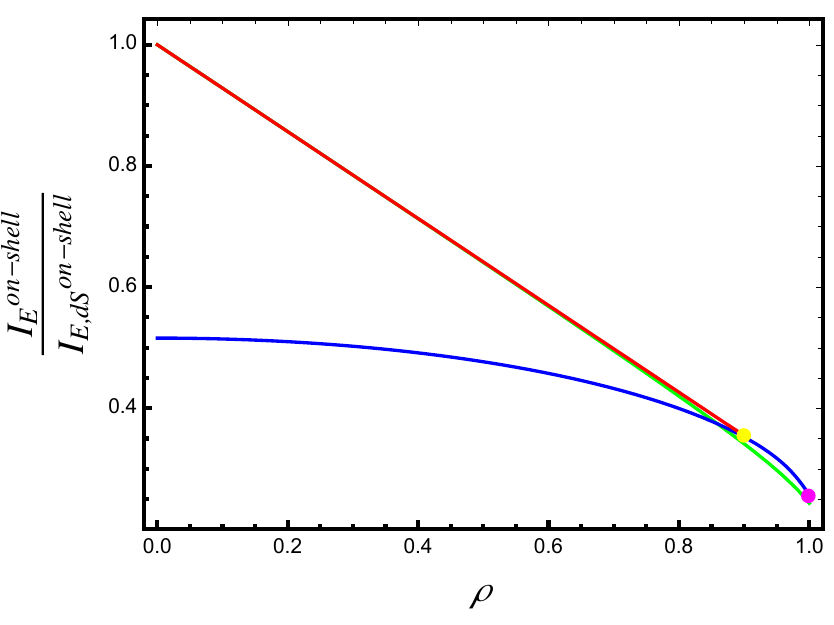}
    \end{minipage}
    \hfill
   \caption{\small Fraction of the Euclidean action $I_{E}^{\text{on-shell}}$ over the action of pure de Sitter $I_{E,dS}^{\text{on-shell}}$ as a function of the mass parameter $M/M_U$ (\textbf{left}) and charge parameter $\rho$ (\textbf{right}) for $\nu=1/3$. The blue line corresponds to the quantum charged Nariai black hole, the green line to the quantum extremal solution, and the red line to the quantum lukewarm solution. The (lower) magenta dot is the ultracold limit and the (upper) yellow dot the Nariai lukewarm solution.}
  \label{fig:I}
\end{figure}

In Figure \ref{fig:I} we display the total on-shell Euclidean actions for the extremal, Nariai, lukewarm, and ultracold solutions. 
Our results are in qualitative agreement with the classical computation \cite{Morvan:2022aon}.\footnote{Notably, the results of \cite{Morvan:2022aon} differ from those found by Mann and Ross \cite{Mann:1995vb}. Specifically, the on-shell actions for the extremal and ultracold solutions differ as Mann and Ross, following \cite{Hawking:1994ii}, take the extremal black hole to have zero entropy.} We  note the actions of the lukewarm (red curve) and extremal (green curve) solutions coincide at $M/M_U = \rho=0$,
and for the ultracold case, we see the extremal and the Nariai (blue curve) solution coincide. 


The goal now is to evaluate the path integral (\ref{eq:nucratearbv2}) using a saddle-point approximation, i.e., evaluating the Euclidean action $I_{E}$ about its 
 (smooth) stationary points -- instantons. As with the classical RN-dS black hole, to ensure the Euclidean charged quantum de-Sitter black hole (cqSdS)  be a stationary point of the Euclidean action, its mass and charge must be fixed via constraints. In this sense, Euclidean cqSdS  is a ``constrained instanton'', that is, a regular solution to the Euclidean field equations after a specific constraint is implemented. At the level of the Euclidean path integral (\ref{eq:nucratearbv2}), this is accomplished by integrating over a pair of Lagrange multipliers, resulting in (see \cite{Morvan:2022ybp,Morvan:2022aon} for details)
\begin{equation}
    \Gamma =  \int d\rho \, d \alpha e^{-(I_{\text{cqSdS}}-I_{\text{qdS}})}= \int d\rho \, d \alpha \, e^{-\Delta S}\;,
\label{eq:nucrateqbhgen}\end{equation}
where $\alpha$ and $\rho$ are, respectively, the dimensionless mass and charge parameters introduced in (\ref{alpha_rho}). To arrive to the second equality we used the the on-shell relation (\ref{eq:Ionshell3d}) and $\Delta S=S^{(3)}_{\text{gen},\text{qdS}}-S^{(3)}_{\text{gen},\text{cqSdS}}$, with $S^{(3)}_{\text{gen,cqSdS}}=S_{\text{gen},+}^{(3)}+S_{\text{gen},c}^{(3)}$ being the total entropy of the charged quantum dS black hole.

\begin{figure}[t!]
  \centering
  \begin{minipage}[t]{0.47\textwidth}
    \includegraphics[width=.9\textwidth]{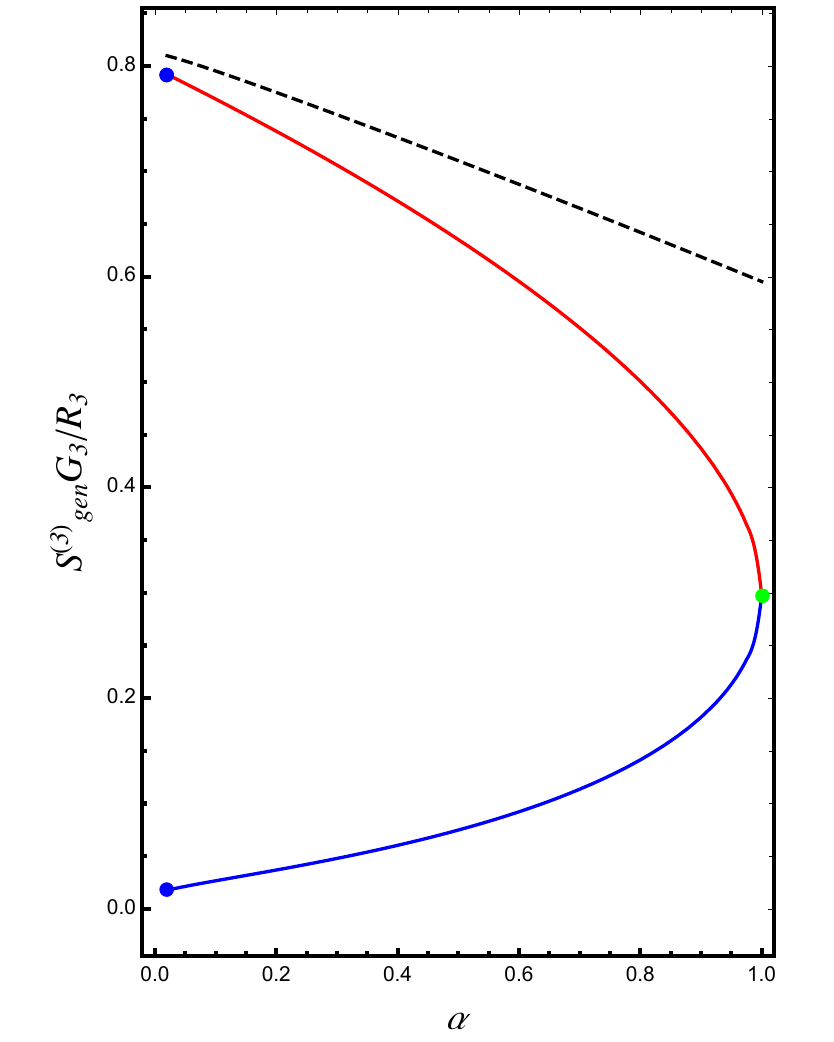}
  \end{minipage}
  \hfill
  \begin{minipage}[t]{0.47\textwidth}
    \includegraphics[width=.9\textwidth]{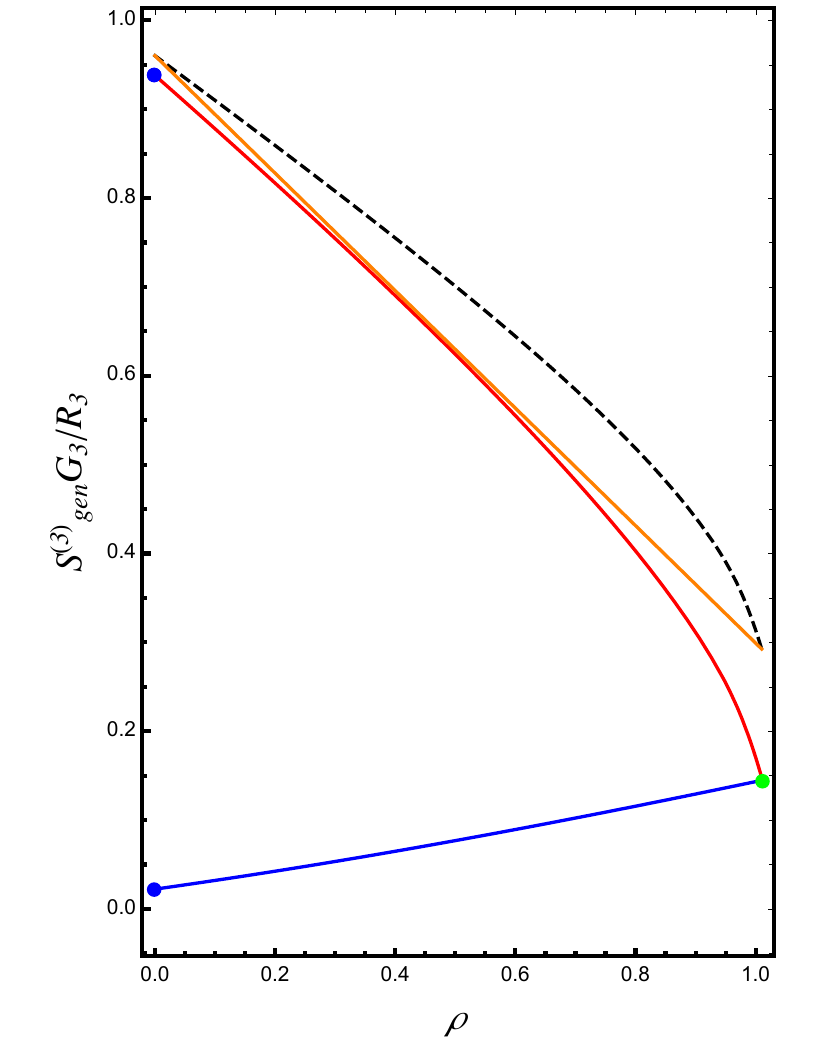}
    \end{minipage}
    \hfill
   \caption{\small  \textbf{Left:} Generalized entropy $S_{\text{gen}}^{(3)}$ as a function of $\alpha$ at $\rho = 0.5$ and $\nu = 1/3$. Blue and red curves correspond to temperatures $S_{\text{gen,}+}^{(3)}$ and $S_{\text{gen,c}}^{(3)}$, respectively. The dashed black line denotes the total generalized entropy, i.e., the sum of black hole and cosmological horizon entropies. The green and blue dots correspond to the Nariai and extremal limit, respectively. \textbf{Right:} Generalized entropy as a function of the dimensionless charge parameter $\rho$ at $\alpha = 0.5$ and $\nu = 1/3$. The orange line shows the linear approximation \eqref{lineal_fit_rho}.}
  \label{fig:S}
\end{figure}

\subsection*{Approximate evaluation of $\Gamma$}

Analytically solving the integral (\ref{eq:nucrateqbhgen}) generically is not possible. We can then proceed via either numerics or, as in~\cite{Morvan:2022ybp,Morvan:2022aon}, by approximating the integrand. A discussion on both approaches for charged black holes in dS$_4$ is provided in Appendix \ref{app:nucleation}. 
As shown in the left panel of Figure \ref{fig:S}, the total entropy of the quantum charged black hole (qcSdS) can be approximated by a linear fit in $\alpha$ at fixed $\rho$,
\beq S_{\text{gen,cSdS}}^{(3)}\approx \alpha S_{\text{gen,N}}^{(3)} + (1-\alpha)S_{\text{gen,E}}^{(3)}\, , \label{eq:linealpha}\eeq
where $S_{\text{gen,N}}^{(3)}$ and $S_{\text{gen,E}}^{(3)}$ are the generalized entropy of the Nariai and extremal black hole, respectively.
Using this approximation, the probability $\Gamma$ for nucleating a quantum dS black hole of arbitrary mass within the sector of fixed $\rho$ is 
\beq \Gamma\approx \int_{0}^{1}d\rho \Gamma_{\rho}\, ,  \quad \text{with} \quad \Gamma_{\rho}= e^{-(S_{\text{gen,qdS}}^{(3)}-S_{\text{gen,E}}^{(3)})}\left(\frac{1-e^{S_{\text{gen,N}}^{(3)}-S_{\text{gen,E}}^{(3)}}}{S_{\text{gen,E}}^{(3)}-S_{\text{gen,N}}^{(3)}}\right)\;.\label{eq:nucrateintegralrho}\eeq
Following a similar approach, we evaluate the integral over $\rho$ by applying the following linear approximation for the extremal and Nariai entropies,
\begin{equation} \label{lineal_fit_rho}
    S_{\text{gen,E}}^{(3)} \approx \rho S_{\text{gen,U}}^{(3)} + \left( 1-\rho \right)S_{\text{gen,dS}}^{(3)} \, , \quad  S_{\text{gen,N}}^{(3)} \approx \rho S_{\text{gen,U}}^{(3)} + \left( 1-\rho \right)S_{\text{gen,N}}^{(3),0} \, .
\end{equation}
where $S_{\text{gen,U}}^{(3)}$ is the entropy of the ultracold black hole and $S_{\text{gen,N}}^{(3),0}$ is the entropy of the neutral $(\rho=0)$ Nariai black hole.
This approximation is illustrated in the right panel of Figure \ref{fig:S}, where the black dashed line is the total generalized entropy, and the orange line shows the linear fit. 

 Using the linear approximations (\ref{eq:linealpha}) and (\ref{lineal_fit_rho}), the nucleation probability of a charged quantum black holes of arbitrary mass and charge is 
\begin{equation}\label{eqn:charging_manus}
    \Gamma \approx \dfrac{e^{-(S_{\text{gen,dS}}^{(3)}-S_{\text{gen,U}}^{(3)})}}{S_{\text{gen,dS}}^{(3)}-S_{\text{gen,N}}^{(3),0}} \bigg( \gamma(\sigma_{\text{gen,N}})-\gamma(\sigma_{\text{gen,dS}}) + \log{(\sigma_{\text{gen,N}}/\sigma_{\text{gen,dS}})} \bigg) \, ,
\end{equation}
where $\gamma(\sigma)$ is the incomplete Euler function (Eq. (\ref{eq:incompEufunc})), and $\sigma_{\text{N}}\equiv S_{\text{gen,U}}^{(3)}-S_{\text{gen,N}}^{(3),0}$ and $\sigma_{\text{dS}}\equiv S_{\text{gen,U}}^{(3)}- S_{\text{gen,dS}}^{(3)}$. 


\subsection*{Numerical evaluation of $\Gamma$}

Another method for calculating the nucleation rate involves numerical techniques, as previously discussed. The most interesting case for a numerical evaluation is the full nucleation probability, obtained by integrating over both charge and mass parameters. In part, this is because the approximate form of the probability~\eqref{eqn:charging_manus} still requires numerical evaluation of the Euler function to study it in detail. While numerical evaluation of the full nucleation probability is conceptually straightforward, a direct implementation encounters significant slow downs due to the determination of $(\rho, \alpha)$. While manageable in the evaluation of $\Gamma_\rho$, this becomes prohibitive in the double integral. To circumvent this issue, our numerical implementation makes use of a change of integration variables from $(\alpha, \rho)$ to $(a, p)$ defined as
\be
p = \frac{q}{q_{\rm U}} \, , \quad \mu = a \mu_{\rm N} + (1-a) \mu_{\rm E} \, .
\ee
In the above, $q_{\rm U}$ denotes the parameter $q$ for the ultracold solution, while $\mu_{{\rm N}/{\rm E}}$ denote the parameter $\mu$ for the Nariai and extremal solutions. After this change of variables, the resulting integral is
\be 
\Gamma = \int_0^1 dp \int_0^1 da \left| \frac{\partial(\mu, q)}{\partial(a, p)} \right| \left| \frac{\partial(\alpha, \rho)}{\partial(\mu, q)} \right| e^{-\Delta S} \, ,
\ee
where $\partial(\alpha, \rho)/\partial(\mu, q)$ and $\partial(\mu, q)/\partial(a, p)$ denote the Jacobian of the transformations. All derivatives appearing in the Jacobians can be evaluated analytically and reduced to expressions in which the only undetermined parameter is $x_1$. The analysis, while straightforward, results in expressions that are rather complicated and therefore we do not present them here. The final integral, while requiring this manual preprocessing, is relatively quick and efficient to evaluate. 

We use our numerical implementation to study how the nucleation probability depends on the backreaction parameter $\nu$. Before presenting the result, let us first make a few important remarks. Varying the backreaction parameter $\nu$ in general corresponds to varying a combination of the parameters of the theory, $G_3, R_3$ and the central charge of the CFT, $c$. The precise relationship is given by 
\be\label{G3R3} 
\frac{c G_3}{R_3} = \frac{\nu}{2 \sqrt{1-\nu^2}} \, .
\ee
Depending on which quantities among $(G_3, R_3, c)$ are held fixed gives rise to different interpretations of the corresponding variations in $\nu$. 

\begin{figure}[t!]
    \centering
    \includegraphics[width=0.75\linewidth]{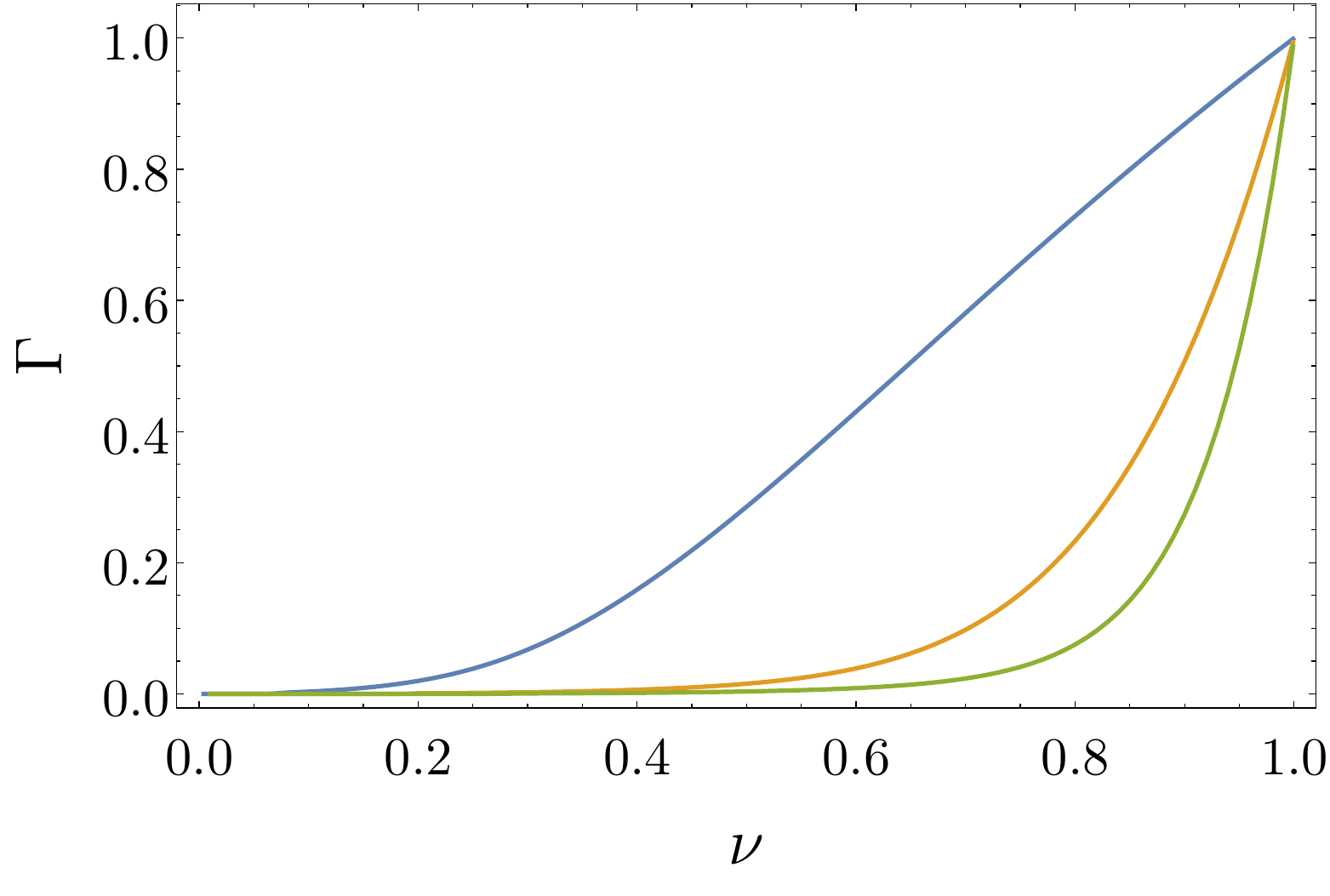}
    \caption{Nucleation probability for $c = 1, 5, 10$ corresponding to the (from left to right) blue, orange and green curves, respectively. Here we have set $R_3 =1$. The limit $\nu \to 0$ corresponds to $G_3 \to 0$, while $\nu \to 1$ corresponds to $G_3 \to \infty$.}
    \label{fig:full_nuc}
\end{figure}

Here we will study what is perhaps the most natural interpretation, which is to fix $c$ and $R_3$, while allowing $G_3$ to vary. In this way, when $\nu \to 0$ we have $G_3 \to 0$ and gravity on the brane decouples.  We present the nucleation probability as a function of $\nu$ at fixed central charge and $R_3$ in Figure~\ref{fig:full_nuc} for three different values of the central charge. We observe that as $\nu \to 0$, corresponding to $G_3 \to 0$, the nucleation probability goes to zero. The nucleation probability approaches unity as $\nu \to 1$, corresponding to $G_3 \to \infty$. This behavior matches qualitatively with the nucleation probability for a conical deficit in dS$_3$ --- see appendix~\ref{app:nucleation}.

Increasing the central charge has the effect of suppressing the nucleation probability across a larger portion of the parameter space. This can be seen in the fact that the orange and green curves hug the horizontal axis more tightly than the blue curve.

\section{Outlook}\label{sec:disc}

Using braneworld holography, we constructed the first charged three-dimensional de Sitter black hole exactly incorporating all orders of semi-classical backreaction. Similar to the uncharged and rotating quantum dS$_{3}$ black holes, the black hole horizon appears due to backreaction of a large-$c$ holographic CFT$_{3}$ with a cutoff in the UV. That the black hole is charged is a consequence of the backreacting CFT being charged; in the absence of backreaction the quantum black hole reduces to the neutral Schwarzschild-dS$_{3}$ solution with a single cosmological horizon. As the dS$_{3}$ black hole is charged, it has extremal, ultracold, and charged Nariai limits, with a parameter space governed by an analog of the shark-fin diagram characteristic of higher-dimensional classical RN-dS black holes.


Analyzing the horizon thermodynamics, we found the classical entropy is replaced by the semi-classical generalized entropy in the first law. Further, we have demonstrated that the heat capacity of quantum dS$_3$ black holes exhibit (inverted) Schottky peaks analogous to their higher-dimensional classical Schwarzschild- or Reissner-Nordstr{\"o}m-de Sitter counterparts. Specifically, for a specific finite range of temperature, both classical and quantum charged black holes behave as thermal systems with a finite number of energy levels available to its underlying microscopic degrees of freedom, beyond which many energy levels become available. This indicates charged dS black holes have a large gap in their energy spectrum.

Finally, we used (constrained) instanton methods to compute the probability to nucleate charged quantum dS$_{3}$ black holes of arbitrary mass and charge. The probability for fluctuation from empty dS$_{3}$ to an excited state is primarily controlled by the deficit of generalized entropies between (quantum) dS$_{3}$ and the ultracold and Nariai black holes. We computed the rate/probability in two ways: semi-analytically, using a pair of linear approximations suitable for the analog computation of classical dS$_{4}$ black holes, and exactly using numerics. We found the nucleation probability, at fixed central charge, decreases as the strength of backreaction increases, i.e., when gravity becomes non-dynamical.

There are a number of avenues worth exploring, some of which we detail below.

\subsection*{Schottky peaks for quantum black holes}

Here we demonstrated the heat-capacity of quantum dS$_3$ black holes exhibit Schottky peaks analogous to their higher-dimensional Schwarzschild- or Reissner-Nordstr{\"o}m-de Sitter counterparts. This is consistent with the fact that dS black holes have a finite mass range, bounded above by the Nariai mass. Notably, all three-dimensional anti-de Sitter quantum black holes have a finite range of mass. As such AdS$_{3}$ quantum black holes are expected to feature Schottky behavior (indeed, in an ensemble of fixed cosmological constant, the heat capacity of the static quantum BTZ black hole \cite{Frassino:2023wpc,Johnson:2023dtf} has inverted peaks). This is in stark contrast with classical anti-de Sitter black holes, which only exhibit Schottky behavior in ensembles of fixed (thermodynamic) volume \cite{Johnson:2019vqf}. It would be worth seeing if asymptotically flat quantum black holes \cite{Emparan:2002px,Panella:2024sor} also feature Schottky behavior. Doing so would establish that the appearance of Schottky peaks, at least in three-dimensions, are a universal quantum signature.

\subsection*{Quasi-normal modes of quantum dS black holes} 

In this article we focused on static quantum black holes. One way to explore the dynamics of quantum black holes is to consider time-dependent perturbations to static or stationary black holes, and study characteristic traits encoded in their quasi-normal mode (QNMs) spectrum. The QNMs of spin-0 and spin-$1/2$ fields were analyzed for the quantum BTZ black hole \cite{Cartwright:2024iwc}, which were subsequently used to characterize the pole-behavior of the correlators of the dual field theory. It would be interesting to extend and adapt this analysis to study the QNM spectrum of quantum dS black holes. The computation of such QNMs would also prove indispensable for tests of strong quantum cosmic censorship \cite{Davey:2024xvd}. 

\vspace{2mm}

\subsection*{Extended thermodynamics of quantum dS black holes}

Here we considered thermal ensembles with a fixed cosmological constant. Alternatively, one could consider the extended framework where the cosmological constant is not fixed but instead treated as dynamical pressure \cite{Kastor:2009wy,Dolan:2010ha}, $P\propto -\Lambda$. In fact, braneworld holography facilitates a natural origin for extended thermodynamics of quantum black holes, where tuning the tension of the end-of-world brane induces a variable cosmological constant on the brane \cite{Frassino:2022zaz}. For dS black holes, where $P<0$, it is arguably more sensible to interpret a variable cosmological constant as a tension, consistent with the higher-dimensional origin via holographic braneworlds. Akin to their AdS counterparts, dS black holes are conjectured to obey a set of reverse isoperimetric inequalities \cite{Dolan:2013ft}, placing an upper bound on their (cosmological) horizon entropy. It would be interesting to study the extended thermodynamics of quantum dS black holes and see if they obey a set of quantum inequalities \cite{Frassino:2024bjg}, or explore their critical phase behavior via thermodynamic geometry as done for the neutral quantum BTZ \cite{HosseiniMansoori:2024bfi}.



\subsection*{Decay of extremal quantum black holes}

The weak gravity conjecture \cite{Arkani-Hamed:2006emk} states, roughly, that large extremal black holes in asymptotically flat or AdS spacetimes should decay into to smaller, sub-extremal black holes, in accordance with (weak) cosmic censorship. Practically, the weak gravity conjecture imposes an upper bound on the mass of an elementary charged particle. The weak gravity conjecture is among a set of consistency conditions characterizing the ``swampland'' paradigm \cite{Ooguri:2006in,Palti:2019pca} to distinguish low energy effective field theories that have a UV completion in quantum gravity from those which do not. More carefully, effective field theories with a $U(1)$ gauge field consistent with quantum gravity should contain a ``superextremal'' massive, charged state (emitted by a black hole via Schwinger pair creation) of mass $m_{s}$ and charge $q_{s}$ that obeys 
\beq q_{s}/m_{s} \gtrsim 1\;,\label{eq:WGCbound}\eeq
where the precise inequality depends on the choice of units and spacetime dimension.\footnote{Due to negative curvature effects of AdS, the charge-to-mass ratio in AdS is modified, see, e.g., \cite{Nakayama:2015hga,Montero:2018fns,Aharony:2021mpc}.} Proofs of the weak gravity conjecture have been offered in \cite{Montero:2018fns,Cheung:2018cwt}, relying on the thermodynamic description of charged black holes.

In a similar vein, the ``Festina-Lente'' conjecture \cite{Montero:2019ekk,Montero:2021otb} asserts a lower bound on the mass of charged particles in a de Sitter black hole background, such that large charged black holes discharge and evaporate to empty de Sitter space. Otherwise, when this mass bound is not observed, charged Nariai black holes discharge rapidly, becoming superextremal Nariai (neutral dS black hole with a mass exceeding the Nariai bound), such that the system evolves towards a big crunch, and violates weak cosmic censorship.\footnote{Notably, however, incorporating backreaction due to the emission of massive, charged shells results in certain decay channels which violate the Festina-Lente bound yet do not result in a big crunch \cite{Aalsma:2023mkz}.} Specifically, preventing the formation of naked singularities during the decay of extremal Nariai demands (in four spacetime dimensions)
\beq m_{s}^{2} \gtrsim M_{\text{P}}H q_{s}\;,\label{eq:FLbound}\eeq
for Hubble constant $H$. The Festina-Lente bound (\ref{eq:FLbound}) is compatible with the weak gravity constraint (\ref{eq:WGCbound}), thus placing an upper and lower bound on $m_{s}$, and leads to a host of phenomenological implications \cite{Montero:2021otb}.  

It is natural to wonder either the weak gravity and Festina-Lente conjectures are robust against quantum matter backreaction. The charged quantum BTZ black hole \cite{Climent:2024nuj} and dS$_{3}$ black hole constructed here provide a precise case study to explore this question. Following the arguments of \cite{Cheung:2018cwt}, it is at least plausible large extremal charged quantum BTZ black holes are unstable to decay to smaller extremal quantum black holes, and the weak gravity conjecture is satisfied. The reasoning relies on the fact that, for perturbatively small backreaction, the infinite tower of higher-derivative corrections in the induced brane action contribute positively to black holes with positive heat capacity. Alone, this implies corrections to the (three-dimensional) Bekenstein-Hawking entropy result in a positive entropy shift, and indicates the bound (\ref{eq:WGCbound}). It remains to be seen precisely how the cutoff CFT affects this entropy shift, requiring further study. 

Testing the perdurance of the Festina-Lente bound in the presence of quantum matter can be done by utilizing the holographic construction of the quantum dS$_{3}$ black hole. In particular, the constraint (\ref{eq:FLbound}) follows from disallowing certain decay channels due to Schwinger pair creation \cite{Montero:2019ekk}, reaffirmed  in \cite{Aalsma:2023mkz} using an $s$-wave tunneling formalism in the probe limit. Adapting these techniques to the AdS$_{4}$ bulk geometry dual to the dS$_{3}$ braneworld would provide the first example of a Festina-Lente type bound for dS$_{3}$ (quantum) black holes.  



\appendix

\section*{Acknowledgements}
We thank Roberto Emparan, Juan Pedraza, and Manus Visser for useful discussions. AC is supported by MICIN through grant PRE2021-098495 and by the State Research Agency of MICIN through the ‘Unit of Excellence Maria de Maeztu 2020-2023’ award to the Institute of Cosmos Sciences (CEX2019-000918-M). The work of RAH is supported by a fellowship from ``la Caixa” Foundation (ID 100010434) and from the European Union’s Horizon 2020 research and innovation programme under the Marie Skłodowska-Curie grant agreement No 847648 under fellowship code LCF/BQ/PI21/11830027. EP is supported by the cosmoparticle initiative at University College London. AS is supported by
STFC grant ST/X000753/1.

\section{Range of Parameters} \label{app:range_bulk}


The metric functions $H(r)$ and $G(x)$ in \eqref{eq:AdS4Ccoord} control the existence of horizons and the transverse sections' geometry, respectively. Consequently, the solutions that can be described on the brane depend on the root structure of $G(x)$. We characterize these solutions as the smallest root of $x$, namely $x_1$, and the parameter $q$, as stated in \eqref{eq:muqtdef}.

The black hole horizons are described by the roots of $H(r)$. Only three of the four real roots (assuming there are any) will be positive. Assuming $\mu>0$ and $q$ fixed, if we gradually increase the value of $\mu$ from $0$ there is only one root: the cosmological horizon. Consequently, there is a naked singularity as we are above the extremal limit. This continues until $\mu$ reaches $\mu_E$, the extremal value for which $H(r)$ has only two roots. At this stage, the outer and inner horizons coincide, while the cosmological horizon, which is larger than the extremal horizon radius, is still present. As we continue increasing $\mu$ beyond $\mu_E$, we enter a regime where there are three positive roots: the inner, outer and cosmological horizon. This continues until $\mu$ reaches the Nariai limit $\mu_N$, where there are once again only two roots, but this time the outer and the cosmological horizons coincide. When $\mu>\mu_N$, there are no longer any real roots. Consequently, the extremal black hole represents the smallest black hole that can fit inside the static patch, whereas the Nariai black hole is the largest black hole to fit inside the static patch.

To obtain the explicit expressions of $\mu_E$ and $\mu_N$ as a function of $q$ and $\nu$ \eqref{eq:defznu}, we solve $H(r)=H'(r)=0$, resulting in
\be \label{eq:muEmuN}
    \mu_E = \sqrt{\dfrac{2}{3}}\dfrac{12q^2\nu^2-\sqrt{1-12q^2\nu^2}}{3\nu\sqrt{1-\sqrt{1-12q^2\nu^2}}} \, , \quad \mu_N = \sqrt{\dfrac{2}{3}}\dfrac{1+12q^2\nu^2+\sqrt{1-12q^2\nu^2}}{3\nu\sqrt{1+\sqrt{1-12q^2\nu^2}}} \, .
\ee
In the limit in which $q$ is small we find
\be
    \mu_E = 2q+\mathcal{O}(q^2) \, , \quad \mu_N = \dfrac{2}{3\sqrt{3}\nu} +\mathcal{O}(q^2) \, .
\ee
We see $\mu_E$ vanishes as $q\rightarrow 0$ while $\mu_N$ remains non-zero. This is what we would expect since, in this limit, extremal black holes do not exist. Additionally, any black hole cannot be arbitrarily large; it must be confined within the cosmological horizon. Notice that both extremal and Nariai black holes exist only if 
\be
    q < \dfrac{1}{2\sqrt{3}\nu} \, .
\ee

Using \eqref{eq:muEmuN}, we can obtain the horizon radius $r_0$ for each limit 
\be
\label{eq:radii}
    (r_0)_E = \dfrac{R_3}{\sqrt{6}}\sqrt{1-\sqrt{1-12q^2\nu^2}} \, , \quad (r_0)_N = \dfrac{R_3}{\sqrt{6}}\sqrt{1+\sqrt{1-12q^2\nu^2}} \, .
\ee
We can relate the mass and charge of the extremal/Nariai limits by combining \eqref{eq:muEmuN} and \eqref{eq:radii}.  After a little algebra, 
\begin{equation}
    \mu_{E/N} = 2 \left(1-2 \frac{(r_0)^2_{E/N}}{R_3^2} \right) \frac{(r_0)_{E/N}}{R_3} \frac{1}{\nu} \ , \qquad |q_{E/N}| = \left(1-3 \frac{(r_0)^2_{E/N}}{R_3^2} \right)^{1/2} \frac{(r_0)_{E/N}}{R_3} \frac{1}{\nu} \ ,
\end{equation}
where, recall, $r_E \in [0,r_u]$ and $r_N \in [r_u, R_3/\sqrt{3}]$.

Using our expressions for $\mu$, we can solve maximum and minimum values for the $x_{1}$ parameter (the smallest positive real root of $G(x)=0$). In particular, substituting $\mu = \mu_E$ into \eqref{eq:muqtdef} we get the maximum value of $x_1$, denoted as $x^{\rm max}_1$. Meanwhile, setting $\mu = \mu_N$ we find the minimum value $x^{\rm min}_1$. While explicit expressions are cumbersome, we can write down perturbative expansions in terms of small $q$ and $\nu$:
\beq
    x^{\rm max}_1 = 1-q+2q^2 + \mathcal{O}(q^3) \, ,
\eeq
\beq
    x^{\rm min}_1 = \dfrac{\sqrt{3}}{2^{1/3}}\nu^{1/3} + \mathcal{O}(q^2) \, ,
\eeq
which we confirm are positive.  Notice we recover the appropriate neutral qSdS$_{3}$ result as $q \rightarrow 0$. Specifically,  $x_{1}^{\text{max}}=1$ and $x_{1}^{\text{min}}=x_{1}^{\text{N}}$ (see Eq. (5.14) of \cite{Emparan:2022ijy}).




\section{Nucleation of classical black holes and defects}  \label{app:nucleation}

\subsection*{Nucleation of charged dS$_4$ black holes}

Following \cite{Morvan:2022ybp,Morvan:2022aon}, the nucleation probability $\Gamma$ of a classical electrically charged dS$_4$ black hole \eqref{eq:RN_dS} of any given mass and charge is
\begin{equation}
    \Gamma =  \int d\rho \, d \alpha \, e^{-\left( S_{\text{dS}} - S_{\rm RNdS}\right)}\;.
\label{eq:nucrateqbhgenAPP}\end{equation}
Here $S_{\text{dS}}$ is the entropy of empty dS$_4$ and $S_{\rm RNdS}$ is the total entropy of a charged black hole in dS$_4$, i.e., the sum of the Bekenstein-Hawking entropies of the black hole and the cosmological horizon. The dimensionless parameters $\alpha$ and $\rho$ are related to the black hole mass and charge via \eqref{alpha_rho}. This integral cannot be solved analytically in general. In \cite{Morvan:2022ybp,Morvan:2022aon}, a pair of linear approximations were made to proceed. Specifically, the total entropy can be approximated using a linear fit in $\alpha$ at fixed $\rho$,
\beq S_{\text{RNdS}} \approx \alpha S_{\text{N}} + (1-\alpha)S_{\text{E}}\, , \label{eq:linealphaAPP}\eeq
where $S_{\text{N}}$ and $S_{\text{E}}$ are the entropy of the Nariai and extremal black hole, respectively. 
With this approximation, the probability $\Gamma$ \eqref{eq:nucrateintegralrhoAPP} to nucleate an arbitrary mass dS black hole in a sector of fixed $\rho$ is 
\beq \Gamma\approx \int_{0}^{1}d\rho \Gamma_{\rho} \, ,  \quad \text{with} \quad \Gamma_{\rho}= e^{-(S_{\text{dS}}-S_{\text{E}})}\left(\frac{1-e^{S_{\text{N}}-S_{\text{E}}}}{S_{\text{E}}-S_{\text{N}}}\right)\; .\label{eq:nucrateintegralrhoAPP}\eeq
To evaluate the integral over $\rho$, further use the linear approximations for the extremal and Nariai entropies,
\begin{equation} \label{eq:lineal_fit_rhoAPP}
    S_{\text{E}} \approx \rho S_{\text{U}} + \left( 1-\rho \right)S_{\text{dS}} \, , \quad  S_{\text{N}} \approx \rho S_{\text{U}} + \left( 1-\rho \right)S_{\text{N}}^{0} \, .
\end{equation}
where $S_{\text{U}}$ and $S_{\text{N}}^{0}$ are the entropy of the ultracold and the neutral $(\rho=0)$ Nariai black hole, respectively. With this approximation, the probability of nucleation is
\begin{equation}
    \Gamma \approx \dfrac{e^{-(S_{\text{dS}}-S_{U})}}{S_{\text{dS}}-S_{N}^{0}} \bigg( \gamma(\sigma_{N})-\gamma(\sigma_{\text{dS}}) + \log{(\sigma_{N}/\sigma_{\text{dS}})} \bigg) \, ,
\label{eq:nucrateintAPP}\end{equation}
where $\sigma_{N}\equiv S_{U}-S_{N}^{0}$, $\sigma_{\text{dS}}\equiv S_{U}- S_{\text{dS}}$, and $\gamma(\sigma)$ is the Euler gamma function
\beq
    \gamma(\sigma)\equiv \int_{\sigma}^{\infty} t^{-1} e^{-t} \, dt\,.
\label{eq:incompEufunc}\eeq

\begin{figure}[t!]
  \centering
  \begin{minipage}[t]{0.47\textwidth}
    \includegraphics[width=.9\textwidth]{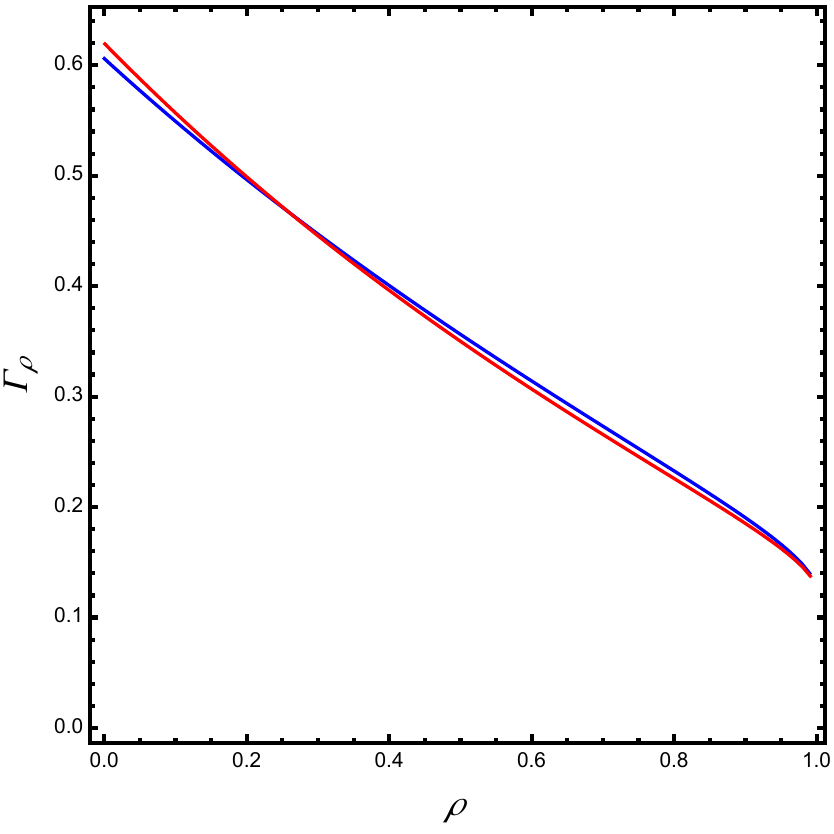}
  \end{minipage}
  \hfill
  \begin{minipage}[t]{0.47\textwidth}
    \includegraphics[width=.9\textwidth]{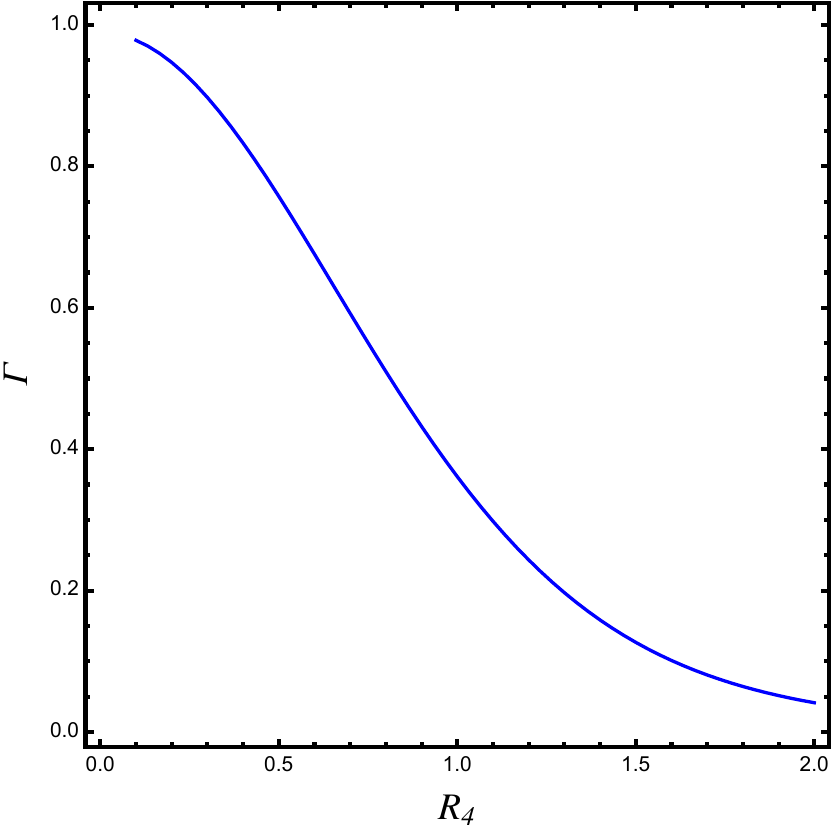}
    \end{minipage}
    \hfill
   \caption{\small  \textbf{Left:} Nucleation probability of a charged dS$_4$ black hole in a sector at fixed $\rho$, i.e. $\Gamma_{\rho}$, with constant dS length $R_{4}=1$. The blue curve corresponds to the numeric result and the red is the linear approximation \eqref{eq:nucrateintegralrhoAPP}. \textbf{Right:} Nucleation probability of charged dS$_4$ black hole as a function of the dS length $R_{4}$ obtained by numerical methods.}
  \label{fig:ClassicalGamma}
\end{figure}

To check the validity of the above approximations, we evaluated the nucleation rate using numerical methods. The left plot of Figure \ref{fig:ClassicalGamma} shows \eqref{eq:nucrateintegralrhoAPP} as a function of $\rho$ for both methods, the numerical result (blue curve) and the linear approximation (red curve).
We further evaluated \eqref{eq:nucrateqbhgenAPP} numerically (right panel of Figure \ref{fig:ClassicalGamma}) where we find the nucleation probability decreases as the de Sitter radius increases.

\subsection*{Nucleation of neutral defects in dS$_{3}$}

Unlike in higher dimensions, we can exactly analytically compute the probability to nucleate a conical defect in empty dS$_{3}$. To this end, recall the Schwarzschild-dS$_{3}$ metric
\beq ds^{2}=-f(r)dt^{2}+f^{-1}(r)dr^{2}+r^{2}d\phi^{2}\;,\quad f(r)=1-8G_{3}M-\frac{r^{2}}{R_{3}^{2}}\;.\eeq
When $8G_{3}M<1$, the geometry describes a conical defect with a cosmological horizon $r_{c}=R_{3}\sqrt{1-8G_{3}M}$, and deficit angle $2\pi(1-\sqrt{1-8G_{3}M})$; for $8G_{3}M>1$, the defect geometry has no horizon and an excess (surfeit) angle of $2\pi(\sqrt{1-8G_{3}M}-1)$. Note $M=1/8G_{3}\equiv M_{\text{N}}$ sets an upper bound on the point mass generating the defect. In this (singular) limit the conical geometry has a $2\pi$ deficit and can be understood as a having a big-crunch or big-bang singularity \cite{Balasubramanian:2001nb}. The Gibbons-Hawking entropy is 
\beq S_{\text{SdS}_{3}}=\frac{\pi R_{3}}{2G_{3}}\sqrt{1-8G_{3}M}\;,\eeq
which reduces to the Gibbons-Hawking entropy for empty dS$_{3}$ when $M=0$ and vanishes when $M=M_{\text{N}}$. 

Applying the formalism developed in \cite{Morvan:2022ybp} for arbitrary spacetime dimension, we can exactly compute the probability $\Gamma$ to nucleate a defect in dS$_{3}$ of arbitrary mass. Specifically, 
\beq 
\begin{split}
 \Gamma=\int_{0}^{1}d\alpha e^{-\Delta S}&=\frac{4G_{3}}{\pi R_{3}}\left[\frac{2G_{3}}{\pi R_{3}}\left(e^{-\pi R_{3}/2G_{3}}-1\right)+1\right]\\
  &=\frac{2}{S_{\text{dS}_{3}}}\left[\frac{1}{S_{\text{dS}_{3}}}\left(e^{-S_{\text{dS}_{3}}}-1\right)+1\right]\;,\label{eq:exactrate}
\end{split}
\eeq
for entropy deficit $\Delta S=S_{\text{dS}_{3}}-S_{\text{SdS}_{3}}$, and $\alpha=M/M_{\text{N}}$. We observe $\Gamma\to1$ as the entropy $S_{\text{dS}_{3}}\to0$, while $\Gamma\to0$ as the Gibbons-Hawking entropy becomes infinite. Equivalently, for fixed $R_{3}$, the nucleation probability is maximized in a regime where $G_{3}\to\infty$, and vanishes when gravity is absent, $G_{3}\to0$.

It is worth comparing our exact result to the analogous linear approximation for the entropy deficit
\beq S^{\text{fit}}_{\text{SdS}_{3}}=S_{\text{dS}_{3}}-(S_{\text{dS}_{3}}-S_{\text{N}})\alpha=S_{\text{dS}_{3}}(1-\alpha)\;,\label{eq:SdS3entfit}\eeq
and consequently, $\Delta S\approx -S_{\text{dS}_{3}}\alpha$. The probability is then approximately 
\beq \Gamma_{\text{fit}}=\int_{0}^{1}d\alpha e^{-S_{\text{dS}_{3}}\alpha}=\frac{1}{S_{\text{dS}_{3}}}\left(1-e^{-S_{\text{dS}_{3}}}\right)\;.\label{eq:linfitprob}\eeq
The probability $\Gamma_{\text{fit}}$ coincides with the vanishing backreaction limit of the quantum SdS$_{3}$ black hole (cf. Eq. (6.19) of \cite{Emparan:2022ijy} once appropriately normalized). 

Comparing the exact probability \eqref{eq:exactrate} to one computed using the linear fit \eqref{eq:linfitprob}, we see the linear approximation well captures the exponential decay of the rate, however, the two probabilities significantly differ as $S_{\text{dS}_3}$ becomes large,
\beq \lim_{S_{\text{dS}_{3}}\to\infty}\Gamma\approx 2/S_{\text{dS}_{3}}=2\Gamma^{\infty}_{\text{fit}}\;.\eeq
The reason for this discrepancy is simple. When performing the linear approximation to the action in the spirit of \cite{Morvan:2022ybp}, one is really interpolating the entropy so as to minimize the error over the whole range $\alpha \in [0,1]$, especially at the endpoints. Such an approximation, however, is insufficient due to the exponential weighting in the integral.
By enforcing the error to be zero at $\alpha=1$, there is an enhanced error near $\alpha\approx0$, a crucial portion of the integration domain. 

\begin{figure}[t!]
  \centering
  \begin{minipage}[t]{0.47\textwidth}
    \includegraphics[width=.9\textwidth]{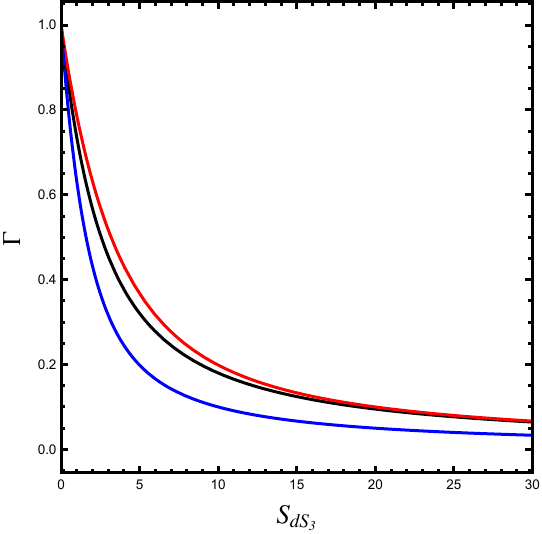}
  \end{minipage}
  \hfill
  \begin{minipage}[t]{0.47\textwidth}
    \includegraphics[width=.9\textwidth]{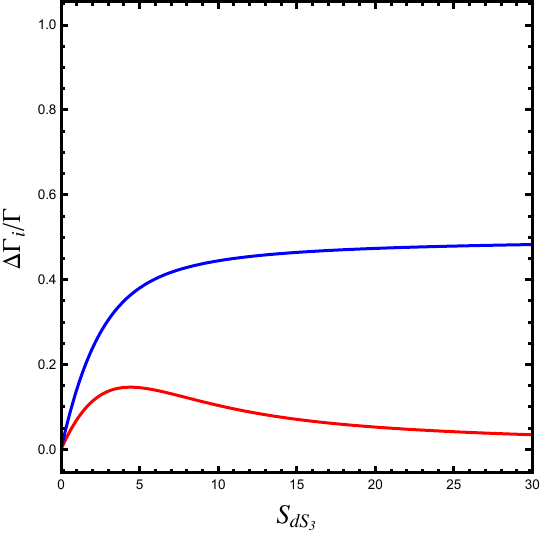}
    \end{minipage}
    \hfill
   \caption{\small  \textbf{Left:} Nucleation probability for the SdS$_3$ geometry as a function of the empty dS$_3$ entropy. The (middle) black line is the exact solution, whilst the (top) red and (bottom) blue curves are the approximations obtained via the expansion around $\alpha=0$ and interpolation (\ref{eq:SdS3entfit}), respectively. \textbf{Right:} Relative error in $\Gamma$ for the two approximations with respect to the exact solution.
   }
  \label{fig:approximations_rate}
\end{figure}

To see this, consider a Taylor expansion of the entropy deficit $\Delta S(\alpha)$ about small $\alpha$, 
\begin{equation}
    \Delta S(\alpha) \approx S_{\text{dS}_3} \left(1-\frac{\alpha}{2} \right) \ ,
\end{equation}
where we keep only terms linear in $\alpha$. 
Substituting this into the probability $\Gamma$ and performing the integration gives
\begin{equation}
    \Gamma_{\text{Tay}} =\frac{2}{S_{\text{dS}_3}} \left(1-e^{-S_{\text{dS}_3}/2} \right) \ .
\end{equation}
This approximation is a closer fit to the exact numerical result, compared to $\Gamma_{\text{fit}}$, as we illustrate in Figure \ref{fig:approximations_rate}. Thus, while true that the linear interpolation (\ref{eq:SdS3entfit}) better approximates the entropy deficit \cite{Morvan:2022ybp} (especially as the spacetime dimension is increased), at least for $d=3$, we see the fit using a small $\alpha$ expansion of the deficit better approximates the nucleation probability.


\bibliography{qdSrefs}

\end{document}